\documentclass[acmsmall,screen]{acmart}
\usepackage{tabularx}  
\usepackage{xcolor}
\usepackage{graphicx}
\usepackage{subfigure}
\usepackage{subcaption}
\usepackage{longtable}
\usepackage{pdflscape}

\usepackage[pagewise]{lineno}
 
\AtBeginDocument{%
  }

\setcopyright{acmlicensed}
\copyrightyear{2018}
\acmYear{2018}
\acmDOI{XXXXXXX.XXXXXXX}

\acmJournal{JACM}
\acmVolume{37}
\acmNumber{4}
\acmArticle{111}
\acmMonth{8}

    \usepackage{color}
    \usepackage{xcolor}
    
    \newcounter{ChenhuiCommentCounter}
\newcounter{RubingCommentCounter}
\newcounter{AbelCommentCounter}
\begin{document}

\title{A Survey on Web Application Testing: Over a Decade of Evolution}

\author{Tao Li}
\email{3220007015@student.must.edu.mo}
\orcid{0009-0001-7413-9692}
\affiliation{
  \institution{School of Computer Science and Engineering, Macau University of Science and Technology}
  \city{Macao SAR}
  \postcode{999078}
  \country{China}  
}
\affiliation{
  \institution{School of Artificial Intelligence (Institute of Artificial Intelligence), Beihang University}
  \city{Beijing}
  \postcode{100191}
  \country{China}
}

\author{Chenhui Cui}
\email{3230002105@student.must.edu.mo}
\orcid{0009-0004-8746-316X}
\author{Zhengdong Huang}
\email{1230014123@student.must.edu.mo}
\orcid{0009-0007-2488-3898}
\affiliation{
  \institution{School of Computer Science and Engineering, Macau University of Science and Technology}
  \city{Macao SAR}
  \postcode{999078}
  \country{China}
}

\author{Rubing Huang}
\email{rbhuang@must.edu.mo}
\orcid{0000-0002-1769-6126}
\affiliation{
  \institution{School of Computer Science and Engineering, Macau University of Science and Technology}
  \city{Macao SAR}
  \postcode{999078}
  \country{China} 
}
\affiliation{
  \institution{Macau University of Science and Technology Zhuhai MUST Science and Technology Research Institute}
  \city{Zhuhai}
  \state{Guangdong}
  \postcode{519099}
  \country{China}
}

\author{Dave Towey}
\email{dave.towey@nottingham.edu.cn}
\orcid{0000-0003-0877-4353}
\affiliation{
  \institution{School of Computer Science, University of Nottingham Ningbo China}
  \city{Ningbo}
  \state{Zhejiang}
  \postcode{315100}
  \country{China}  
}

\author{Lei Ma}
\email{ma.lei@acm.org}
\orcid{0000-0002-8621-2420}
\affiliation{
  \institution{School of Computer Science, University of Tokyo}
  \city{Tokyo}
  \country{Japan}
  \postcode{113-0033}
}
\affiliation{
  \institution{Department of Electrical and Computer Engineering, University of Alberta}
  \city{Edmonton}
  \country{Canada}
  \postcode{T6G 2R3}
}

\author{Yuan-Fang Li}
\email{yuanfang.li@monash.edu}
\orcid{0000-0003-4651-2821}
\affiliation{
    \institution{Faculty of Information Technology, Monash University}
   \city{Melbourne}
   \state{Victoria}
   \postcode{3800}
   \country{Australia}
}

\author{Wen Xia}
\email{xiawen@hit.edu.cn}
\orcid{0000-0003-4093-6391}
\affiliation{
  \institution{School of Computer Science and Technology, Harbin Institute of Technology Shenzhen}
  \city{Shenzhen}
  \state{Guangdong}
  \postcode{518055}
  \country{China}
  }

\begin{abstract}

    As one of the most popular software applications, a web application is a program accessible through the web that dynamically generates content based on user interactions or contextual data; examples include online shopping platforms, social networking sites, and financial services. 
    Web applications operate in diverse environments and leverage web technologies such as HTML, CSS, JavaScript, and Ajax, often incorporating features like asynchronous operations to enhance user experience.
    Due to the growing number of users and the popularity of web applications, the quality of these applications has become increasingly important. 
    \textit{Web Application Testing} (WAT) plays a vital role in ensuring the functionality, security, and reliability of web applications.
    Given the speed with which web technologies are evolving, WAT is especially important.
    In the last twelve years, various WAT approaches have been developed.
    The diversity of approaches reflects the many aspects of web applications, such as dynamic content, asynchronous operations, and diverse user environments.
    This paper provides a comprehensive overview of the main achievements over the last twelve years: 
    It examines the main steps involved in WAT, including test case generation and execution, as well as evaluation and assessment.
    The currently available tools for WAT are also examined. 
    The paper also discusses open research challenges and potential future work in WAT.

\end{abstract}



\received{20 August 2026}


\maketitle

\section{Introduction
\label{SEC:introduction}}

The \textit{Software Development Life Cycle} (SDLC)~\cite{shafiq2021literature} comprises several essential phases, including requirements specification, architecture design, coding, testing, and maintenance~\cite{hanna2022web}. 
Each phase involves specific processes and activities that contribute to the overall development of the software. 
Among these phases, testing is particularly critical, often accounting for a significant share of the time and cost of developing a software application.
Recent studies~\cite{lopez2022machine,bindu2022survey} indicate that the testing process can consume 10\% to 80\% of the total development time, depending on project complexity.
The primary objective of testing is to identify errors that manifest during program execution. 
An effective testing methodology improves the probability of detecting such errors, for example, by systematically running and examining a wide range of execution paths, including edge cases and fault-prone areas~\cite{li2014two,mahdieh2022test}.

Since the adoption of the World Wide Web in the early 1990s~\cite{BERNERSLEE1992454}, web applications have become integral to various sectors, including finance, healthcare, education, and commerce~\cite{dougan2014web}. 
These applications have introduced innovative features and functionalities that facilitate consistent and efficient access to information across diverse sources, significantly transforming how society interacts with digital content~\cite{jazayeri2007some}. 
Meanwhile, web applications rely on a set of core technologies, including HTML to define the structure of web pages, CSS to style them, JavaScript to enable interactivity, and AJAX for asynchronous data exchange between the client and the server. 
These technologies are the foundation of modern web applications and are critical in supporting their dynamic and distributed nature~\cite{garrett2005ajax,arora2013dynamic,sampath2016advances}. 

As web applications grow increasingly complex and technologies continue to evolve, the requirement for robust and effective testing methodologies has become more crucial~\cite{atkinson2002towards}:
It is necessary to continuously evolve and advance testing strategies to ensure the reliability, security, and overall quality of web applications~\cite{li2014two}.
\textit{Web Application Testing} (WAT)~\cite{dougan2014web} has become an essential process in the SDLC of web applications, and is used to evaluate and ensure the quality of \textit{Applications Under Test} (AUTs)~\cite{karimoddini2022automatic}, making it a cornerstone of modern software quality assurance. 
WAT aims to ensure that web applications function as intended, meet specified requirements, and are free from vulnerabilities. 
The scope of WAT goes beyond detecting and fixing bugs to include enhancing user experience, ensuring data security, and maintaining the overall reliability of web systems. 
As web technologies evolve, testing methodologies have advanced to address the increasing complexity and demands of web applications.

Web applications have become integral to modern society~\cite{garousi2013systematic}, influencing various aspects of daily life and industrial practice~\cite{onukrane2023navigating}.
Accordingly, several reviews have examined specific aspects of WAT.
For example, \citet{prazina2023methods} reviewed the methods for testing and analyzing web layouts.
Web application security has also received considerable research attention~\cite{rahman2022mapping,aydos2022security,zhang2021efficiency,altayaran2021integrating,cui2020survey,wang2020access,muslihi2020detecting,marashdih2019web,kumar2016review,rafique2015web,gupta2014International}, whereas 
\citet{gupta2018automated} specifically examined the automated generation of regression test cases for web applications.
Other studies have provided different perspectives on WAT. For example,
\citet{dougan2014web} conducted a systematic literature review that identified and summarized the major research topics in WAT.
\citet{balsam2024web} focused on the challenges and opportunities in WAT.
More recently, \citet{kertusha2026survey} surveyed studies published between 2014 and 2025, analyzing research trends, testing methodologies and tools, applications of \textit{Artificial Intelligence} (AI), experimental practices, and industrial engagement.
These studies offer valuable but complementary perspectives, differing in their thematic scope, temporal coverage, and analytical focus.

The earlier survey~\cite{li2014two} led by Yuan-Fang Li
---
who is also a co-author of this paper
---
is closely related to the present study.
That survey mainly reviewed two decades of WAT research, covering testing techniques, research challenges, and developments in the field up to 2013.
By comparison, the present study focuses on WAT research published from 2014 onward, providing a complementary examination of more recent developments.
Specifically, this paper adopts a comprehensive, life-cycle-oriented perspective on WAT.
It synthesizes the current state of research on test-case generation and execution, evaluation metrics and methods, and supporting tools.
Failure diagnosis and regression testing are excluded as standalone research topics; however, failure-detection capability is considered as a metric for evaluating test effectiveness.
Furthermore, it identifies key open challenges and outlines promising directions for future WAT research.

The rest of this paper is organized as follows:
Section~\ref{SEC:background} provides background information on the formal definition of WAT and the fundamental testing techniques employed in the field. 
Section~\ref{SEC:methodology} outlines the survey methodology, including the six research questions, the literature-retrieval process, and the statistical analysis of the retrieved studies.
Sections~\ref{SEC:rq1} to~\ref{SEC:rq6} address each of the research questions (RQ1 to RQ6), respectively. 
Finally, Section~\ref{SEC:conclusions} concludes the paper by summarizing key findings and proposing directions for future research.

\section{Background
\label{SEC:background}}

This section presents preliminary WAT concepts and provides an overview of web applications and WAT, including core definitions and relevant testing techniques.

\subsection{Web Applications}

A web application~\cite{jazayeri2007some} is a software system that facilitates user interaction through a combination of front-end and back-end components, over a network, typically accessed through a web browser. 
The front-end, or client-side, renders the user interface and handles user inputs. 
It consists of web pages constructed using technologies such as HTML, CSS, and JavaScript.
This layer is designed to provide a responsive and interactive user experience, enabling seamless communication between the user and the application. 
The back-end, or server-side, encompasses the core functionalities of the application:
These include data processing, business logic execution, and database management. 
The back-end is the key to the application:
It processes front-end requests and ensures that the correct data and functionality are provided to the user. 
Communication between the front-end and back-end typically relies on protocols such as HTTP and HTTPS, which support smooth data exchange and maintain the application’s state~\cite{kumar2017security,neumann2018analysis}.
Together, these components provide an integrated system that supports a wide range of user interactions, from simple content browsing to complex transaction processing.

Web applications can be either static or dynamic. 
Static web applications have fixed content that remains unchanged regardless of user interactions, making them suitable for delivering straightforward information, such as in digital brochures or basic company websites~\cite{vaidyanathan2009security}. 
Dynamic web applications, in contrast, provide a more interactive experience, with content that adapts to user inputs, interactions, or real-time data~\cite{sampath2016advances}. 
The dynamic nature of these applications enables greater flexibility and user engagement, making them well-suited for complex platforms such as social media networks, e-commerce sites, and online service portals that require continuous interaction and real-time responsiveness~\cite{fulcini2022gamified}.

\subsection{Web Application Testing}

Web applications have become a fundamental part of our daily lives, making thorough testing a vital process to ensure their functionality and dependability.
Effective testing ensures that these applications not only meet the required functionality but also satisfy security, performance, and reliability requirements. 
Thorough testing of web applications can help to identify and address potential vulnerabilities, performance bottlenecks, and functional issues:
This enhances the user experience and helps to safeguard sensitive data, which is vital for maintaining trust in web-based services.
Comprehensive testing ultimately contributes to the development of robust, dependable web applications that can adapt to the dynamic demands of users and the evolving technological landscape.

WAT involves several key steps:
\begin{itemize}
    \item 
    Analysis of the functional and non-functional requirements of the web application to guide the development of comprehensive test cases.
    
    \item 
    Generation of test data according to the testing methodology, such as model-based or security testing, and design of appropriate test cases to ensure broad coverage of potential scenarios.
    
    \item 
    Execution of the test cases by sending HTTP requests to the web application, and comparison of the responses with the expected outcomes to identify inconsistencies.
    
    \item 
    Documentation of the test results, analysis of any issues, and application of regression testing to ensure that recent changes have not introduced new defects.
    
    \item 
    Implementation of continuous testing and monitoring to maintain the application’s stability and security as it evolves.
\end{itemize}

\section{Methodology
\label{SEC:methodology}}

In this section, we present the six research questions (RQs) related to WAT in our study.
Our WAT survey employed a systematic, structured methodology guided by the published frameworks of \citet{huang2019survey} and \citet{webster2002analyzing}.
Our presentation of findings draws on recent surveys on related topics, including testing of RESTful APIs~\cite{golmohammadi2023testing}, mutation analysis~\cite{jia2010analysis}, and metamorphic testing~\cite{segura2016survey,chen2018metamorphic}. 
The following sections detail key aspects of the review process and results.

\subsection{Research Questions}

The goal of this survey paper was to identify and categorize the available WAT information. 
To achieve this, the following six RQs guided the study:

\begin{itemize}
    \item 
    \textbf{RQ1:} 
    \textit{How has WAT evolved, and what WAT topics have been examined in published studies?}
    
    \item
    \textbf{RQ2:} 
    \textit{What methods and approaches have been used for WAT test case generation?}
    
    \item 
    \textbf{RQ3:} 
    \textit{How are test cases executed in WAT?}

    \item 
    \textbf{RQ4:} 
    \textit{What metrics are used to evaluate WAT? How are they used?}
    
    \item 
    \textbf{RQ5:} 
    \textit{What tools are available to support WAT?}
    
    \item 
    \textbf{RQ6:} 
    \textit{What are the challenges and future work related to WAT?}
\end{itemize}

\begin{figure}[!t]
  \centering
  \includegraphics[width=\textwidth]{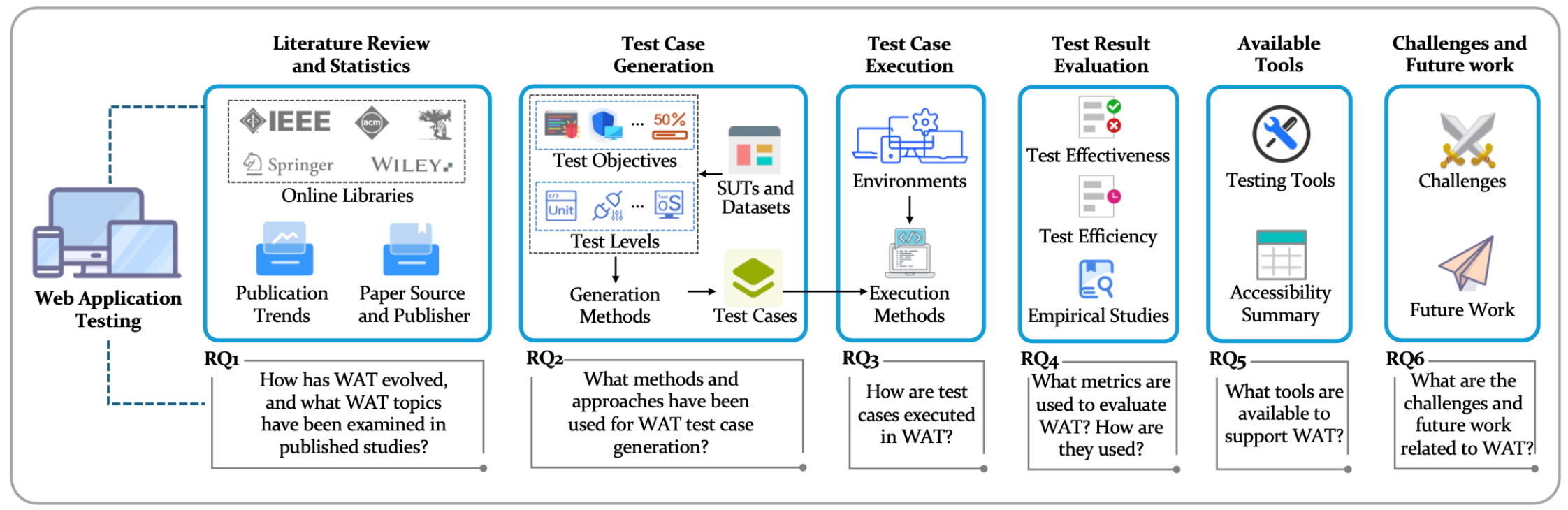 }
  \caption{Structure of this survey paper.}
  \label{fig:workflow}
\end{figure}

RQ1 provides an overview of the WAT literature and the distribution of its key topics.
RQ2 examines the methods and approaches used for WAT test-case generation.
RQ3 focuses on the execution of WAT test cases and clarifies the associated processes and practices.
RQ4 identifies the metrics and evaluation frameworks used to assess WAT effectiveness and efficiency.
RQ5 catalogs the currently available WAT tools and provides an overview of the technological resources supporting WAT.
Finally, based on the findings for the preceding RQs, RQ6 identifies unresolved challenges and proposes directions for future WAT research.
Figure~\ref{fig:workflow} presents an overview of the structure of this paper.

\subsection{Literature Search and Selection}

Following previous survey studies~\cite{huang2019survey,webster2002analyzing,golmohammadi2023testing,harman2015achievements,anand2013orchestrated}, we selected the following online repositories of technical research literature:

\begin{itemize}
    \item ACM Digital Library (\textit{ACM})
    \item Elsevier Science Direct (\textit{Elsevier})
    \item IEEE Xplore Digital Library (\textit{IEEE})
    \item Springer Online Library (\textit{Springer})
    \item Wiley Online Library (\textit{Wiley})
\end{itemize}

These repositories were chosen for their extensive collections of conference, symposium, and workshop papers, as well as for providing access to multiple journals considered highly relevant to WAT~\cite{huang2019survey}. 
Our survey examined computer science papers published between January 1, 2014, and December 31, 2025.

Once the literature repositories were identified, search strategies were developed and tailored for each of them, using WAT-specific terminology and formatting. 
Given the unique characteristics and limitations of the advanced search functions in each repository (such as Elsevier’s restriction on using a maximum of eight Boolean connectors per field), the search methods were designed to strictly adhere to the repository rules and constraints. 
To ensure comprehensive coverage of all relevant literature, multidimensional and multilayer keyword filters were used to maximize retrieval of WAT materials. 
Table~\ref{TAB:selected-digital-library-search-queries} illustrates the specific search queries for five selected digital libraries.

\begin{table*}[!t]
\scriptsize
\caption{Selected Digital Libraries, with Search Queries}
\label{TAB:selected-digital-library-search-queries}
    \begin{tabularx}{\textwidth}{c|c|m{11.5cm}}
        \hline
        \textbf{No.} & \textbf{Name} & \textbf{Search Query} \\ \hline
        1 & ACM & 
        \texttt{[[Title: ``web''] OR [Title: ``browser based'']] AND} 
        \texttt{[[Title: ``test''] OR [Title: ``testing''] OR} 
        \texttt{[Title: ``measure''] OR [Title: ``measurement''] OR} 
        \texttt{[Title: ``measuring''] OR [Title: ``check''] OR [Title: ``checking''] OR} 
        \texttt{[Title: ``detect''] OR [Title: ``detecting''] OR [Title: ``detection'']]} \\ \hline
        2 & Elsevier & 
        \texttt{((``web'') OR (``browser based'')) AND}
        \texttt{((``test'') OR (``testing'') OR  (``measure'') OR} 
        \texttt{(``check'') OR  (``checking'') OR} 
        \texttt{(``detect'') OR (``detecting''))} \\ \hline
        3 & IEEE & 
        \texttt{(``Document Title'':``web'' OR ``Document Title'':``browser based'') AND} 
        \texttt{(``Document Title'':``test'' OR ``Document Title'':``testing'' OR} 
        \texttt{``Document Title'':``measure'' OR ``Document Title'':``measurement'' OR} 
        \texttt{``Document Title'':``measuring'' OR ``Document Title'':``check'' OR} 
        \texttt{``Document Title'':``checking'' OR ``Document Title'':``detect'' OR} 
        \texttt{``Document Title'':``detecting'' OR ``Document Title'':``detection'')} 
         \\ \hline
        4 & Springer & 
        \texttt{``web'' AND (``test'' OR ``testing'' OR} 
        \texttt{``measure'' OR ``measurement'' OR ``measuring'' OR} 
        \texttt{``check'' OR ``checking'' OR}  
        \texttt{``detect'' OR ``detecting'' OR ``detection'')} 
         \\ \hline
        5 & Wiley &
        \texttt{``(``web'') OR (``browser based'')'' in Title and} 
        \texttt{``(``test'') OR (``testing'') OR} 
        \texttt{(``measure'') OR (``measurement'') OR (``measuring'') OR}  
        \texttt{(``check'') OR (``checking'') OR} 
        \texttt{(``detect'') OR (``detecting'') OR}  
        \texttt{(``detection'')'' in Title}  
         \\ \hline
    \end{tabularx}
\end{table*}

The survey included English-language studies focused on WAT but excluded academic theses (e.g., master's and doctoral dissertations), books, posters, and prior survey studies. 
To ensure that the selected studies provided sufficient methodological and experimental detail for analysis, we further excluded papers shorter than five pages in IEEE double-column format or seven pages in LNCS single-column format, following the practice adopted in prior survey research~\cite{nguyen2015extensive}.
In addition, some processes relevant to WAT, such as failure diagnosis and regression testing for web applications, were excluded from the survey.
The inclusion criteria were:

\begin{itemize}
    \item The paper was written in English.
    \item The paper was related to WAT.
    \item The paper was not a thesis.
    \item The paper was not a survey or a systematic literature review.
    \item The paper was freely available in open access.
\end{itemize}

After duplicates were removed and the exclusion criteria applied, the initial set of candidate studies was significantly refined. 
A snowballing approach~\cite{golmohammadi2023testing}, which involved examining the reference lists of the selected studies, led to the identification and inclusion of several additional papers. 
Table~\ref{tab:selection_result} summarizes the details of the search and filtering process. 
Eventually, 464 papers were included for the preliminary review and statistical analysis.
To facilitate transparency and reproducibility, the complete list of the 464 papers is publicly available in our accompanying GitHub repository at \url{https://github.com/abelli1024/wat-survey}.

Although some papers may have been omitted due to the focus on a subset of reputable publishers, we are confident that the overall trends we report on are accurate and provide a fair representation of the current state of the art in WAT.

\begin{table*}[!b]
  \caption{Selection Results of Primary Studies}
  \centering
  \scriptsize
  \label{tab:selection_result}
  \setlength{\tabcolsep}{3.5mm}{
    \begin{tabular}{c|c|c|c|c} \hline
  \textbf{No.} & \textbf{Name} 
 & \textbf{\begin{tabular}[c]{@{}c@{}}No. of studies retrieved\\ by the keyword search\end{tabular}} 

  & \textbf{\begin{tabular}[c]{@{}c@{}}No. of studies excluded\\ based on the selection criteria\end{tabular}}  
  & \textbf{\begin{tabular}[c]{@{}c@{}}No. of studies after\\ the selection criteria\end{tabular}}
  \\\hline
  1 & ACM & 888  & 771 & 117 \\\hline
  2 & Elsevier& 433 & 422 & 11 \\\hline
  3 & IEEE& 1335 & 1068 & 267 \\\hline
  4 & Springer & 4530 & 4470 & 60  \\\hline
  5 & Wiley & 151 & 142 & 9  \\\hline
  \multicolumn{2}{c|}{\textbf{\textit{Total}}}&7337 &6873	&464	\\\hline
  \end{tabular}  
  }          
\end{table*}

\subsection{Data Extraction and Collection}

Key information was gathered from each study on the research motivation, contributions to the field, details of empirical evaluations, common misconceptions, and remaining challenges in WAT. 
The extracted data were reviewed and verified by the survey's co-authors, minimizing the risk of overlooking critical information and reducing potential errors in the analyses.

We organized the RQ data collection methods as follows:

\begin{itemize}
    \item 
    \textbf{RQ1:} 
    \textit{Fundamental information for each paper, such as publication year, type of paper, and paper source.}
    
    \item 
    \textbf{RQ2:} 
    \textit{AUTs, testing objectives, and test-case generation techniques.}
    
    \item 
    \textbf{RQ3:} 
    \textit{Execution methods, tools and platforms, environments, tested items, dataset URL, and open-source status.}
    
    \item 
    \textbf{RQ4:} 
    \textit{Evaluation metrics, methods and approaches, and tools and techniques.}
    
    \item 
    \textbf{RQ5:} 
    \textit{Tool name, type, main features and capabilities, URL, and open-source status.}
    
    \item 
    \textbf{RQ6:} 
    \textit{Overview of the remaining challenges.}
\end{itemize}

\section{ANSWER TO RQ1: Publication Evolution and Distribution
\label{SEC:rq1}}

This section addresses RQ1 through a detailed analysis of publication trends and their distribution across the literature sources. 
Our methodology provides a comprehensive analysis of the evolution and distribution of WAT research over the past twelve years.

\subsection{Publication Trends}

\begin{figure*}[!ht]
\graphicspath{{figures/}}
    \centering
    \subfigure[Number of publications per year.]
    {
        \includegraphics[width=0.48\textwidth]{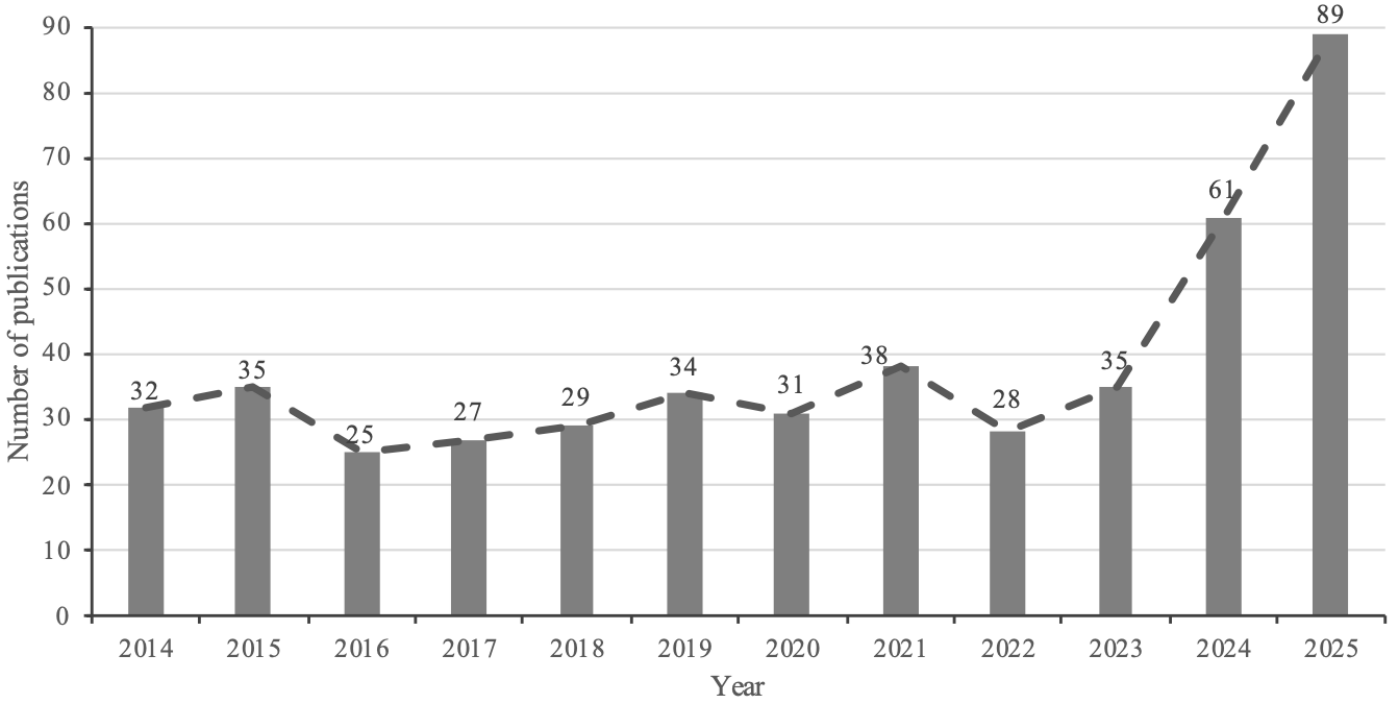}
        \label{Fig:number_of_publications_per_year}
    }
    \subfigure[Cumulative number of publications per year.]
    {
        \includegraphics[width=0.48\textwidth]{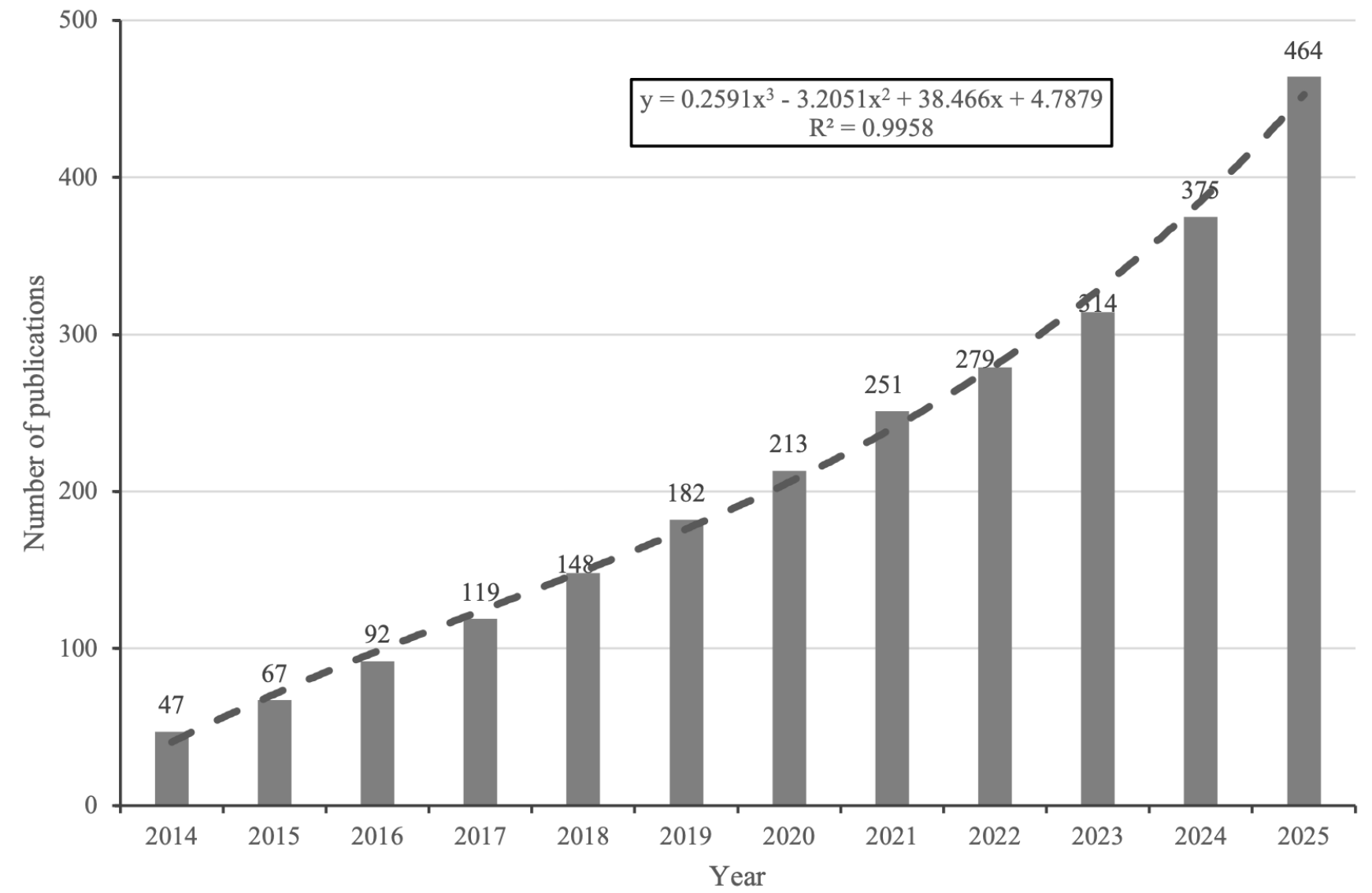}
        \label{Fig:cumulative_number_of_publications_per_year}
    }
    \caption{WAT papers published between January 1, 2014, and December 31, 2025.}
    \label{Fig:publication_trends}
 \end{figure*}

We collected publication year data from 464 papers (as illustrated in Figure~\ref{Fig:publication_trends}) to highlight trends in WAT research from 2014 to 2025. 

Figure~\ref{Fig:number_of_publications_per_year} shows the annual number of publications, and Figure~\ref{Fig:cumulative_number_of_publications_per_year} shows the cumulative total.
Figure~\ref{Fig:number_of_publications_per_year} shows the fluctuating (but relatively steady) nature of WAT-related publications from 2014 to 2025. 
The annual publication counts exhibit some variation over a dozen years, with 32 publications in 2014, followed by slight fluctuations: 
35 in 2015, 25 in 2016, 27 in 2017, 29 in 2018, and 34 in 2019. 
In 2020, the count was 31, which increased to 38 in 2021. 
This was followed by a decrease to 28 in 2022 and a subsequent rise to 35 in 2023. 
In 2024, the count increased to 61, followed by a further increase to 89 in 2025.

The publication output remained relatively stable from 2014 to 2023, followed by substantial growth in 2024 and 2025.
The cumulative data in Figure~\ref{Fig:cumulative_number_of_publications_per_year} indicate sustained growth in WAT research, with a particularly marked acceleration during the final two years of the review period.

\subsection{Paper Source and Publisher}

\begin{figure*}[!t]
\graphicspath{{figures/}}
    \centering
    \subfigure[Proportions of journals and conferences.]
    {
        \includegraphics[width=0.48\textwidth]{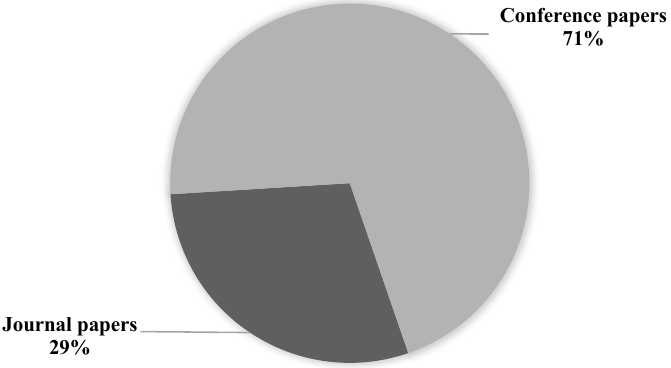}
        \label{Fig:proportion_of_journals_and_conferences}
    }
    \subfigure[Venue distribution per year.]
    {
        \includegraphics[width=0.48\textwidth]{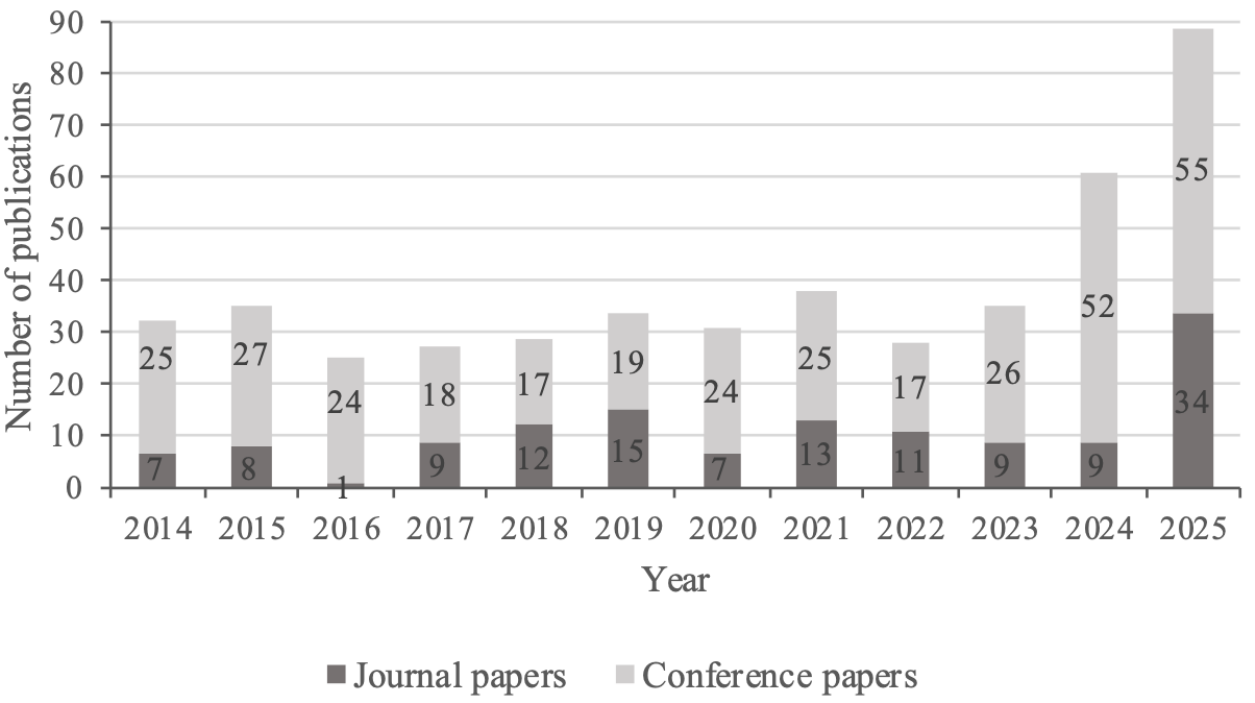}
        \label{Fig:venues_distribution_per_year}
    }
    \caption{Venue distribution of surveyed papers.}
    \label{Fig:venue_distribution_for_papers}
\end{figure*}

The papers analyzed in this study were published in various journals and conference proceedings.
As shown in the table, the majority of the publications were from conferences (71\%), with a smaller proportion being from journals (29\%).     
The data presented in Figure~\ref{Fig:venues_distribution_per_year} reveal distinct patterns in publication volumes over the years. 
For instance, conference papers fluctuated, starting at 25 in 2014, rising slightly to 27 in 2015, then decreasing to 17 in 2018 before recovering to 26 in 2023. 
Conference publication output then rose sharply to 52 papers in 2024, followed by a modest further increase to 55 papers in 2025.
In contrast, journal publications, while also fluctuating, reached at 15 papers in 2019, followed by 9 papers in 2024 and a notable increase to 34 papers in 2025, indicating a recent increase in journal contributions to WAT research.

\section{ANSWER TO RQ2: Test Case Generation
\label{SEC:rq2}}

This section addresses RQ2 by systematically examining key aspects of WAT test case generation, including the \textit{Applications Under Test} (AUTs) and datasets, testing objectives, and test case generation methods. 

\subsection{AUTs and Datasets}

\subsubsection{AUTs}
This section reviews the types of AUTs used in WAT and references the selected research papers.
In WAT, AUTs refer to the web platforms or systems being evaluated (for functionality, performance, security, and usability). 
The diversity of WAT AUTs emphasizes the need for tailored testing approaches to address specific challenges across different domains:
E-commerce platforms, for example, require stringent security measures, whereas government portals may focus on accessibility.

\begin{itemize}
    \item 
    \textbf{\textit{E-commerce Platforms:}} 
    E-commerce applications such as Magento, Shopify, and custom-built online stores are popular WAT AUTs.
    This may be due to their complex transactional workflows, which involve user authentication, shopping carts, and payment gateways~\cite{pratama2022penetration}.
    Testing often focuses on functional correctness and security, with particular attention to vulnerabilities to potential attacks like \textit{SQL Injection} (SQLi) and \textit{Cross-Site Scripting} (XSS)~\cite{bozic2015attack}.
    Performance testing is also essential to ensure that the platforms can handle high traffic volumes during peak times~\cite{wan2019pathmarker}.
    In addition, e-commerce applications such as eBay and ProShop have been adopted in end-to-end testing studies to assess the ability of automatically generated browser tests to exercise realistic shopping workflows~\cite{leotta2024ai}.
    
    \item 
    \textbf{\textit{Content Management Systems (CMSs):}} 
    CMS platforms such as WordPress, Drupal, and Joomla are frequently chosen as AUTs because of their modular and extensible nature, which can make them vulnerable to third-party plugins~\cite{garn2014applicability,arcuri2023emb}.
    Testing typically emphasizes plugin vulnerabilities, core stability, and post-update performance after updates~\cite{li2014automated}.
    Recent WAT research has also employed WordPress as an AUT to investigate the effectiveness of fuzzing-based techniques in detecting vulnerabilities and achieving code coverage in complex web applications~\cite{chen2024webfuzzauto}.

    \item 
    \textbf{\textit{Government and Public Service Portals:}} 
    Government web applications, such as university and public service portals, are often tested for accessibility, usability, and security~\cite{offutt2019testing}.
    These platforms often need to comply with accessibility standards such as the \textit{Web Content Accessibility Guidelines} (WCAG)~\cite{ara2024accessibility}, making accessibility testing critical~\cite{vithanage2016webguardia}. 
    Security testing emphasizes preventing unauthorized access and data breaches, as illustrated by frameworks that target GDPR compliance in web portals~\cite{amanzholova2021development}.
    This category further encompasses healthcare websites, public-sector webpages, disaster-response organization websites, and online service portals.
    Studies of these systems have examined accessibility violations, content reflow, keyboard accessibility, usability, and web performance~\cite{fathallah2025accessguru,chiou2024automatically,iannuzzi2025combined,fischer2025coverage}.

    \item 
    \textbf{\textit{Social Media Platforms:}}
    Social media platforms, like Facebook and Twitter, pose unique challenges for WAT due to their real-time interaction and dynamic content.
    Testing for these applications often examines performance under heavy user loads, cross-browser compatibility, and session management vulnerabilities~\cite{fan2023comprehensive,appelt2018machine}.
    Related online video-sharing platforms have also been used as AUTs in studies assessing web accessibility~\cite{michtner2025web}.
    
    \item 
    \textbf{\textit{Web Application Firewalls (WAFs):}} 
    WAFs play a key role in defending web applications from attacks like SQLi, XSS, and \textit{Cross-Site Request Forgery} (CSRF)~\cite{pelivani2021comparative,zhang2015novel}.
    Testing WAFs involves simulating real-world attack scenarios to evaluate how effectively they distinguish between legitimate and malicious traffic~\cite{zhao2015new,chang2023reinforcement}.
    Recent studies have further evaluated WAFs using generated or mutated HTTP payloads and application-layer traffic traces that emulate injection, XSS, CSRF, and command-injection attacks~\cite{fakhfakh2024combining,tadhani2024securing,wang2024detecting}.    

    \item
    \textbf{\textit{Enterprise Web Applications and Services:}}
    Enterprise systems
    ---
    including single sign-on portals, human resource management systems, point-of-sale systems, industrial web applications, web APIs, and microservice-based systems
    ---
    have also been employed as AUTs in WAT research~\cite{aryotejo2025comparative,shokeh2025comparative,garousi2024coverage,abideen2025intent,wang2024leveraging}.
    Studies involving these systems commonly examine functional and end-to-end (E2E) behavior, automated GUI exploration, model-based coverage, synchronization and waiting strategies, performance, and security properties related to access control, privacy, and APIs~\cite{zhang2024towards,liu2025detecting,feng2025jauthguard,nie2025wascope}.

    \item
    \textbf{\textit{Benchmark Web Applications and Testbeds:}}
    Purpose-built benchmark applications and security testbeds
    ---
    including DVWA, WebGoat, OWASP Juice Shop, WAVSEP, PortSwigger labs, and Vulhub targets
    ---
    are frequently employed as AUTs (or testing targets) because they provide controlled and repeatable environments for evaluating WAT techniques and tools~\cite{bhawsar2025comprehensive,wu2025autopt,lin2024spafuzz,brandi2024sniping,chen2024webfuzzauto}.
    Evaluations using these benchmarks typically assess whether testing techniques and generated inputs can expose known vulnerability classes, such as injection flaws, broken access control, server-side request forgery (SSRF), and request-race vulnerabilities~\cite{liu2025bacscan,chen2025racedb}.

\end{itemize}

\subsubsection{Datasets}
Datasets are essential in WAT, providing the inputs needed to simulate real-world usage and evaluate application behavior across key areas such as security, functionality, and performance. 
The datasets often comprise user data, predefined test cases, or automatically generated data.
They support a range of testing scenarios, including edge cases, and help identify potential vulnerabilities and performance bottlenecks.

Each WAT dataset was designed to meet specific testing objectives, allowing for targeted assessments of various aspects. 
For example, the \textit{Software Assurance Reference Dataset} (SARD)~\cite{fidalgo2020towards} can help to
identify security vulnerabilities, such as SQLi, thus contributing to more secure software development practices. 
The PhishTank dataset~\cite{nirmal2021analyzing}, which contains real-world phishing URLs, can be used to evaluate phishing detection systems and enhance web application security.

Datasets such as the Website Screenshots Dataset~\cite{aditya2023ensemble} facilitate the assessment of visual consistency and UI accuracy, helping ensure an optimal user experience. 
Accessibility-oriented datasets and artifacts
---
including AccessGuru and LangCrUX
---
support the assessment of issues such as WCAG violations and accessibility across multiple languages~\cite{fathallah2025accessguru,bhuiyan2025not}.
Similarly, datasets designed for E2E testing (such as E2E-TestGen) facilitate the evaluation of automated test-generation techniques and their ability to exercise browser-based interactions~\cite{abideen2025intent}.
The NASA web-log dataset~\cite{pavanetto2020generation} provides real-world HTTP requests, offering valuable insights into traffic patterns and server performance under various loads.
DBpedia~\cite{mariani2014link} supports the validation of semantic web applications:
It provides structured data that can help verify data integration and usage.

These datasets enable testers to adopt focused testing strategies to thoroughly evaluate web applications across multiple dimensions.
Table~\ref{TAB:wat_articles} summarizes key datasets used in WAT, with focuses including security, accessibility, performance, and GUI testing.

\begin{scriptsize}
\begin{longtable}{m{2cm}|m{6cm}|m{3.4cm}|c}
\caption{An overview of datasets.} 
\label{TAB:wat_articles} \\
\hline
\textbf{Dataset Name} & \textbf{Overview} & \textbf{Resource Link} & \textbf{Reference} \\ \hline
\endfirsthead

\caption{(Continued).}\\
\hline
\textbf{Dataset Name} & \textbf{Overview} & \textbf{Resource Link} & \textbf{Reference} \\ \hline
\endhead
  Software Assurance Reference Dataset (SARD) 
  & It is a dataset consisting of PHP test cases, where each test case represents a vulnerable or non-vulnerable SQL-related code snippet.
  The test cases include examples of user input propagating through the application code until it reaches a sensitive function (e.g., a database query).
  & \url{https://samate.nist.gov/SARD/index.php}  
  &\cite{fidalgo2020towards}\\\hline
  Near-Duplicate Study DataSet 
  & It is a dataset of about 100,000 annotated pairs of web pages from open-source web applications and real websites. 
  Each pair is labeled as distinct, near-duplicate, or clone to facilitate the evaluation of near-duplicate detection methods.  
  & \url{https://zenodo.org/records/3376730}
  &~\cite{corazza2021web}\\\hline
  Website Screenshots Dataset from Roboflow 
  & It is a dataset of 1,206 screenshots from popular websites with 54,215 UI element annotations, including buttons, text, images, links, titles, fields, iframes, and labels.
  This dataset is used to train object detection models and evaluate the visual consistency of web applications.  
  & \url{https://public.roboflow.com/object-detection/website-screenshots}  &~\cite{aditya2023ensemble}\\\hline
  PhishTank dataset 
  & It is a widely used dataset containing known phishing URLs.
  The dataset provides examples of real phishing attempts that the system uses to test its effectiveness. 
  & \url{http://www.phishtank.com}
  &~\cite{nirmal2021analyzing}\\\hline
  Web injection of publicly available datasets 
  &The datasets consist of various HTTP request samples labeled by the attack type, including SQL injection, XSS, command injection, and path traversal. 
   & SQL Injection: (1) \url{https://www.kaggle.com/syedsaqlainhussain/sql-injection-dataset},
   \newline (2) \url{https://github.com/mhamouei/rat_datasets},
   XSS Injection: \url{https://www.kaggle.com/syedsaqlainhussain/cross-site-scripting-xss-dataset-for-deep-learning},
   HTTP Parameters: \url{https://github.com/Morzeux/HttpParamsDataset}
   &~\cite{romanartificial,amouei2021rat}\\\hline
  Mitch Framework HTTP Request Dataset 
  & It is a dataset based on the Mitch framework that initially contained 5,828 HTTP requests and was reduced to 1,901 records after balancing.
  & \url{https://github.com/alviser/mitch}
  &~\cite{beltrano2023deep}\\\hline
  NASA 1995 Apache Web Log 
  & It is a dataset containing log files from NASA web servers, recording HTTP requests from real users during July 1995.
  & \url{https://ita.ee.lbl.gov/html/contrib/NASA-HTTP.html}
  &~\cite{pavanetto2020generation}\\\hline
  DBpedia 
  & It is a structured knowledge base that extracts information from Wikipedia. 
  It contains millions of entities represented as RDF triples.  
  & \url{https://www.dbpedia.org/}
  &~\cite{mariani2014link}\\\hline
  fuzzdb 
  &It is an open-source database containing numerous attack patterns. 
  & \url{https://github.com/fuzzdb-project/fuzzdb}
  &~\cite{zhang2017anomaly}\\\hline
  ATO-data 
  &It is a collection of keystroke dynamics data collected under controlled lab conditions and anonymized data from 187,332 sessions on financial web pages, which supports analysis of changes in user behavior and is used to evaluate the effectiveness of imposter detection models.  
  & \url{https://github.com/mluckner/ATO-data.git}
  &~\cite{grzenda2023evaluation}\\\hline
  Web Accessibility Evaluation Data 
  & It is an automated web accessibility assessment of nearly 3 million web pages obtained between March and September 2021 using Qual Web.  
  & \url{https://zenodo.org/records/7494722}
  &~\cite{martins2023large}\\\hline
  PHP-Webshell-Dataset
  & It is a dataset containing 2,917 samples from 17 webshell collection projects.  
  & \url{https://github.com/Cyc1e183/PHP-Webshell-Dataset}
  &~\cite{gogoi2022php}\\\hline
  PHP Vulnerability Test Suite
  & It contains 9,408 PHP source-code samples, including 5,600 safe samples and 3,808 unsafe samples.
  The dataset is widely used for detecting XSS vulnerabilities and training predictive models.
  & \url{https://github.com/stivalet/PHP-Vulnerability-test-suite}
  &~\cite{gupta2015text}\\\hline
  Datasets of three common web attack payloads (SQLi, XSS, RCE)
   & It is a dataset generated using an attack grammar and contains SQL-injection, cross-site scripting, and remote-command-execution payloads.
  The dataset is used to train and evaluate GPTFuzzer against WAFs.
  & \url{https://github.com/hongliangliang/gptfuzzer}
  &~\cite{liang2023generative}\\\hline
  Joomla! and MantisBT test suite
  &It is a dataset that contains multiple representative use cases, covering functions such as article and user management. It has 47 manually written test cases and 453 test steps.
  & \url{https://zenodo.org/record/4973219}
  &~\cite{kirinuki2021nlp}\\\hline
  GHTraffic
  & It is a dataset of significant size comprising HTTP transactions extracted from GitHub data (i.e., from the 04 August 2015 GHTorrent issues snapshot) and augmented with synthetic transaction data.
  & \url{https://zenodo.org/record/1034573}
  &~\cite{bhagya2019generating}\\\hline
  AccessGuru Web Accessibility Violations Dataset
  & A benchmark of real web accessibility violations collected from 448 URLs and refined into full, sampled, and semantic human-correction subsets for detection and correction of syntactic, semantic, and layout violations.
  & \url{https://github.com/NadeenAhmad/AccessGuruLLM}
  & \cite{fathallah2025accessguru}\\\hline
  SALAD evaluation subjects
  & A public experiment dataset containing subject webpages and ground truth for detecting reflow accessibility failures in responsive web pages.
  & \url{https://zenodo.org/records/10140605}
  & \cite{chiou2024automatically}\\\hline
  LangCrUX
  & A large-scale multilingual web accessibility dataset used to measure language-related accessibility issues on the web.
  & \url{https://github.com/kal-purush/LangCrux}
  & \cite{bhuiyan2025not}\\\hline
  E2E-TestGen web application benchmark
  & A repository of open-source web applications and injected-fault datasets used to evaluate intent-based E2E test generation and fault-detection ability.
  & \url{https://github.com/E2E-TestGen/Web-Applications.git}
  & \cite{abideen2025intent}\\\hline
  Swedish Public Sector Organizations webpages dataset
  & A public dataset and scripts for evaluating automated accessibility-checking coverage over Swedish public-sector organization webpages.
  & \url{https://doi.org/10.5878/qe0c-kb63}
  & \cite{fischer2025coverage}\\\hline
  WCAG-ACT test cases
  & A public set of W3C accessibility conformance test cases used to evaluate automatic handling of WCAG success criteria.
  & \url{https://github.com/w3c/wcag-act-rules/blob/main/content-assets/wcag-act-rules/testcases.json}
  & \cite{lopez2025turning}\\\hline
  BDD-Cucumber-scenario-Dataset
  & A public set of BDD/Gherkin scenarios used as input for generating Cypress step definitions and page-object-model classes.
  & \url{https://github.com/karpurapus/BDD-Cucumber-scenario-Dataset}
  & \cite{nettur2024cypress}\\\hline
  VETL-data
  & Public experiment data for LVLM-based web GUI testing, including benchmark-web-application materials, commercial-site lists, generated input-text samples, and bug videos.
  & \url{https://github.com/sqlab-sustech/VETL-data}
  & \cite{wang2024leveraging}\\\hline
  Cytestion open-source web applications dataset
  & A dataset of four open-source web applications and associated study artifacts used to compare systematic GUI element discovery approaches.
  & \url{https://gitlab.com/lsi-ufcg/cytestion/loc-study/applications}
  & \cite{de2024automatic}\\\hline
  Ma11y artifacts
  & Public artifacts and repository for mutation-based web accessibility testing, including subject websites, generated accessibility mutants, and tool-analysis resources.
  & \url{https://github.com/mahantaf/web-a11y-tool-analyzer}
  & \cite{tafreshipour2024ma11y}\\\hline
  Higher-education accessibility dataset
  & A Mendeley dataset used to study the impact of COVID-19 on the accessibility of higher-education institution websites.
  & \url{https://data.mendeley.com/datasets/jthx28w7dn/1}
  & \cite{nso2025impact}\\\hline
  Performance-testing online appendix
  & A public replication package for studying adoption, practices, and maintenance of performance testing in open-source web projects.
  & \url{https://figshare.com/s/8d8fc0045e0a627ade0e}
  & \cite{di2025performance}\\\hline
  HTTPParamDataset
  & An open-source HTTP parameter dataset used for command-injection and cross-dataset web attack detection experiments.
  & \url{https://github.com/Morzeux/HttpParamsDataset/}
  & \cite{wang2024detecting}\\\hline
  SQLI--XSS Payload dataset
  & A public Kaggle payload dataset containing normal SQL queries and attack-based SQL queries for SQLi/XSS detection experiments.
  & \url{https://www.kaggle.com/datasets/alextrinity/sqlinjectionextend?select=sqli-extended.csv}
  & \cite{tadhani2024securing}\\\hline
  Kaggle SQL Injection dataset
  & A public SQL Injection dataset with query strings and labels used to train and evaluate SQLi detection models.
  & \url{https://www.kaggle.com/datasets/syedsaqlainhussain/sql-injection-dataset/data}
  & \cite{fakhfakh2024combining}\\\hline
  \texttt{Modified\_SQL\_Dataset.csv}
  & A Kaggle SQL injection dataset containing benign and malicious SQL queries, used for ML-based SQLi detection.
  & \url{https://www.kaggle.com/datasets/sajid576/sql-injection-dataset}
  & \cite{tamilselvan2025sql}\\\hline
  Testcase data set of an MIS system with priority
  & A Kaggle dataset of test-case/task records with requirement priority, weights, function points, complexity, time, cost, and final priority.
  & \url{https://www.kaggle.com/datasets/zumarkhalid/testcase-data-set-of-an-mis-system-with-priority}
  & \cite{manojkumar2024test}\\\hline
  TensorJSFuzz experimental data
  & Public source code and experimental data for testing TensorFlow.js and web-based deep learning APIs via input-constraint extraction.
  & \url{https://sites.google.com/view/gptfjsfuzz}
  & \cite{quan2025tensorjsfuzz}\\\hline
  Industrial web GUI waiting-strategy dataset
  & Public screen recordings and extracted data for 26 business-critical industrial web GUI test cases used to evaluate waiting strategies.
  & \url{https://github.com/senseuwaterloo/ASEIndustryTrack2024}
  & \cite{zhang2024towards}\\\hline
  Free web games ecosystem dataset
  & A public dataset and codebase containing metadata and collected runtime data for approximately 100K free web games.
  & \url{https://github.com/01games/data}
  & \cite{qayyum2024measuring}\\\hline
  Health-related advertising datasets
  & OSF materials for targeted advertising and ad landing-page analysis, including collected ads and manually coded health-related advertising data.
  & \url{https://osf.io/gdx8k/}
  & \cite{zeng2025measuring}\\\hline
  DOM XSS Web Vulnerability Dataset
  & A Carnegie Mellon University dataset used for training web vulnerability detection models focused on DOM XSS vulnerabilities.
  & \url{https://doi.org/10.1184/R1/13870256.v1}
  & \cite{kumar2025web}\\\hline
\end{longtable}
\end{scriptsize}

\subsection{Test Objectives}
WAT test objectives include detecting bugs, ensuring security, evaluating performance, assessing the functionality of \textit{Graphical User Interfaces} (GUIs), and identifying vulnerabilities (such as XSS and SQLi).

\begin{figure}[!b]
    \centering
        \includegraphics[width=0.7\textwidth]        
        {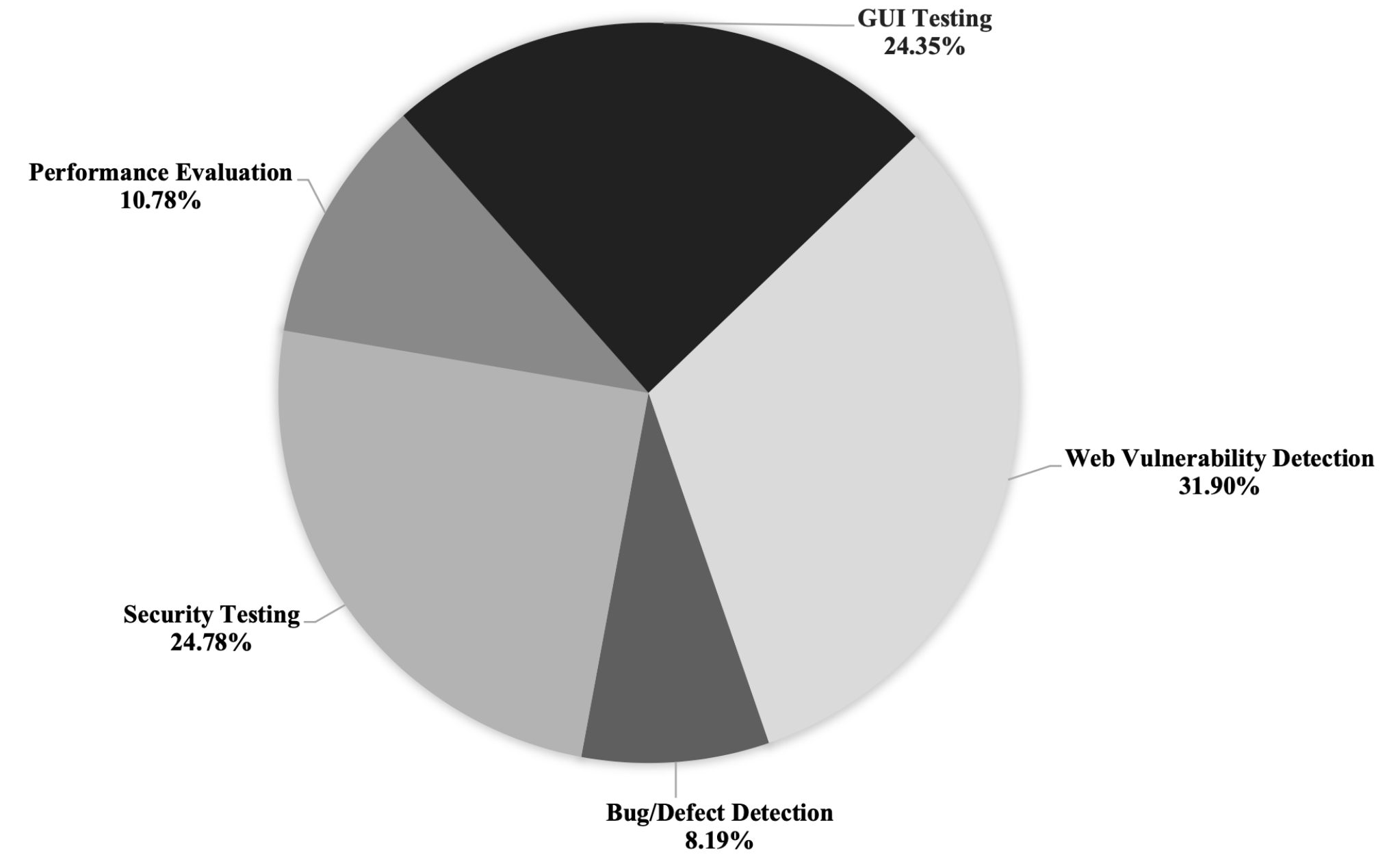}
    \caption{Proportions of test objectives in surveyed papers.}
    \label{Fig:test_objectives_rate}
\end{figure}

Figure~\ref{Fig:test_objectives_rate} shows the distribution of test objectives across the surveyed literature.
Web vulnerability detection, including XSS and SQL Injection detection, accounted for the largest proportion (31.90\%), underscoring the need to address critical security risks. 
Security testing accounted for 24.78\%, underscoring its importance in protecting applications from security threats. 
GUI testing accounted for 24.35\%. Performance evaluation accounted for 10.78\%, and bug/defect detection accounted for 8.19\%.
The following sections provide a detailed analysis of each testing objective.

\subsubsection{Bug/Defect Detection} 
Detecting and correcting defects is essential for maintaining the integrity of web applications.
Techniques such as combining unsupervised learning with embedding layers effectively optimize test case prioritization while reducing redundancy~\cite{demirel2023acum}.
Modern web applications often involve intricate interdependencies between components, requiring advanced techniques to detect defects across diverse states and modules. 
This complexity necessitates strategies such as dynamic test case generation and static analysis to ensure robust coverage~\cite{maskur2019static}.

Static code analysis can complement regression testing and may identify issues early in development. 
Tools that use static analysis (such as taint analysis) detect potential errors in the source code without execution~\cite{maskur2019static,nunes2019empirical}.
Complementary dynamic approaches have also been investigated. 
For example, the Q-learning-based Selenium-Java tool dynamically explores web GUI elements and learns effective test paths, thereby improving test coverage for complex systems~\cite{fan2023comprehensive}.
More broadly, recent studies have extended defect-detection capabilities through automated test-data generation, multi-objective test generation, and the testing of web-based software runtimes and frameworks. 
These approaches are evaluated in terms of the path coverage they achieve and their ability to detect injected faults, incorrect computations, crashes, memory errors, and behavioral inconsistencies across backend implementations~\cite{wanigasekara2024comparing,guo2024multi,quan2025tensorjsfuzz}.

\subsubsection{Security Testing}
The goal of security testing is to ensure that web applications remain protected against unauthorized access, data breaches, and other security threats. 
Security testing provides a systematic framework for evaluating the robustness of applications by identifying and mitigating vulnerabilities before and during runtime~\cite{medeiros2016dekant}.

To achieve comprehensive protection, security testing employs a combination of methodologies. 
Static analysis examines source code for vulnerabilities, such as unhandled exceptions and unsanitized inputs, enabling early detection of issues during development~\cite{zhao2015new}. 
Dynamic analysis evaluates applications at runtime by simulating potential attack scenarios to uncover vulnerabilities that static analysis might overlook~\cite{pellegrino2015jak}. 
Hybrid analysis combines static and dynamic methods to address their individual limitations, offering enhanced coverage and reducing false positives~\cite{guan2021comparative}.
Automated tools, such as Burp Suite, enhance the efficiency of dynamic analysis by enabling real-time evaluation of application defenses~\cite{garn2014applicability}.
Beyond the detection of individual vulnerabilities, security testing also encompasses the orchestration of complete testing workflows and the systematic analysis of privacy risks. 
Automated penetration-testing frameworks coordinate activities such as reconnaissance, vulnerability scanning, exploitation, result aggregation, and report generation. 
In parallel, privacy-focused approaches assess risks associated with browser extensions, web APIs, third-party tracking, and web content that may expose sensitive user data~\cite{wu2025autopt,siddiq2025pungoe,juliadi2025web,nie2025wascope,xie2024arcanum}.

\subsubsection{Performance Evaluation} 
Performance evaluation aims to ensure that an application remains efficient and stable under varying load conditions~\cite{pelivani2021comparative}. 
Essential to this are load and stress testing, which simulate diverse user traffic scenarios to help developers identify performance bottlenecks, especially in high-concurrency environments~\cite{pelivani2021comparative,lounis2014new}. 

Advanced tools such as JMeter~\cite{kiran2015experiences,tiwari2023analytical} and LoadRunner~\cite{cai2014analysis} can automate measurement of some important metrics, including response times, throughput, and system resource utilization (e.g., CPU, memory, disk I/O)~\cite{fan2023comprehensive}.
These measurements can be collected across a wide range of load levels~\cite{fan2023comprehensive}.
These tools can provide comprehensive insights into system behavior under stress, helping to optimize the system architecture and ensure reliability during peak-load conditions~\cite{leotta2020family}.
Beyond conventional load and stress testing, performance evaluation has been investigated in several specialized web contexts:
These include comparisons of web servers under load, spike, and soak workloads~\cite{pinastawa2024web}; 
response-time-guided searches for web API inputs that induce high latency~\cite{huang2024revealing}; 
measurements of front-end energy efficiency~\cite{proma2025energy}; 
maintenance of performance tests in open-source web projects~\cite{di2025performance}; and 
assessments of CI/CD deployment performance in cloud environments~\cite{thokala2025testing}.

\subsubsection{GUI Testing} 
GUI testing aims to ensure that the application’s user interface remains consistent in both functionality and user experience across different devices and browser environments~\cite{zhang2015sampling}. 
Automated tools can simulate user interactions, exploring the GUI’s responsiveness under diverse conditions~\cite{lin2017using}:
Cross-browser testing with tools such as Selenium Grid~\cite{kiranagi2017feature} can help ensure compatibility across platforms~\cite{pelivani2021comparative}.
Meanwhile, Scout uses gamified features~\cite{fulcini2022gamified}, including progress tracking and dynamic error injection, to encourage systematic exploration and improve test coverage.

Image comparison and computer vision technologies are also extensively used to detect interface changes and to identify potential UI errors~\cite{leotta2020family}.
These tools capture and compare screenshots from different stages of testing, ensuring that updates do not introduce unintended changes to the user interface~\cite{zhao2015new}.
The recent integration of AI in testing can further enhance efficiency by learning from user interactions and predicting areas that are likely to fail~\cite{fan2023comprehensive,chang2023reinforcement}.
GUI testing also encompasses accessibility and usability evaluation, as these activities assess whether an interface can be perceived, understood, and operated across diverse user contexts. 
Relevant studies investigate HTML accessibility violations, WCAG-based accessibility mutants, responsive-page reflow, keyboard navigation and focus management, form labeling, color contrast, and interactive-target sizes~\cite{fathallah2025accessguru,tafreshipour2024ma11y,chiou2024automatically,michtner2025web}. 
In parallel, research on GUI automation explores GUI interactions assisted by LVLMs and LLMs, as well as robust synchronization and waiting strategies for E2E tests~\cite{lu2025uxagent,wang2024leveraging,abideen2025intent,zhang2024towards}.

\subsubsection{Web Vulnerability Detection}
Web vulnerability detection focuses on identifying and mitigating specific threats in web applications, such as Cross-Site Request Forgery (CSRF), directory traversal, Cross-Site Scripting (XSS), and SQL Injection (SQLi). 
These vulnerabilities, if left unaddressed, can compromise application functionality, data security, and user privacy~\cite{ben2016grey,zhang2015novel}. 

To detect these vulnerabilities, static analysis examines source code to identify input validation flaws and hardcoded credentials, enabling early detection of coding issues~\cite{zhao2015new,lounis2014new}. 
Dynamic analysis simulates attack scenarios during runtime, evaluating the application’s behavior under unexpected interactions~\cite{ben2016grey}. 
Hybrid analysis combines the strengths of static and dynamic approaches, providing enhanced detection accuracy and reducing false positives~\cite{guan2021comparative}. 

Modern advancements in vulnerability detection integrate machine learning and deep learning techniques. 
Methods based on Recurrent Neural Networks (RNNs) analyze input and output patterns to detect malicious behaviors, demonstrating high effectiveness in identifying XSS and SQLi attacks~\cite{guan2021comparative}. 
Additionally, OWASP ZAP is widely used to automate vulnerability identification, providing robust support for detecting SQLi and XSS~\cite{zhang2015novel}. 
Recent research has broadened this objective to encompass a wider range of security weaknesses and web-based threats:
These include authorization flaws, such as broken access control and missing owner checks; 
request-race conditions; 
recurring vulnerability patterns in PHP applications; 
web cache poisoning; 
webshells and phishing pages; and 
vulnerabilities involving server-side request forgery (SSRF), CSRF, and command injection. 
This breadth demonstrates that modern vulnerability detection extends beyond the classification of injection payloads to the analysis of application-specific authorization logic, concurrent request behavior, cache behavior, malicious code patterns, and security-sensitive API interactions~\cite{feng2025jauthguard,liu2025mocguard,chen2025racedb,shi2024recurscan,liang2024internet,gao2025distilling,liu2025bacscan,ghadiyaram2025phish,manivannan2025ai,ramadan2024enhancing,wang2024detecting}.

\subsection{Generation Methods}
WAT can make use of many test case generation methods, including:
Model-Based~\cite{dias2007survey}, Combinatorial and Optimization-Based~\cite{nie2011survey}, AI-Based~\cite{alagarsamy2024a3test}, Data and Content-Driven~\cite{elbaum2008carving}, Crawling and Exploration-Based~\cite{mesbah2012crawling}, Template and Page Object-Based~\cite{stocco2016automatic}, and Security Test Case Generation techniques~\cite{felderer2016security}. 
These methods offer a broad range of strategies to address diverse testing needs.

\begin{figure}[!t]
    \centering
        \includegraphics[width=0.8\textwidth]{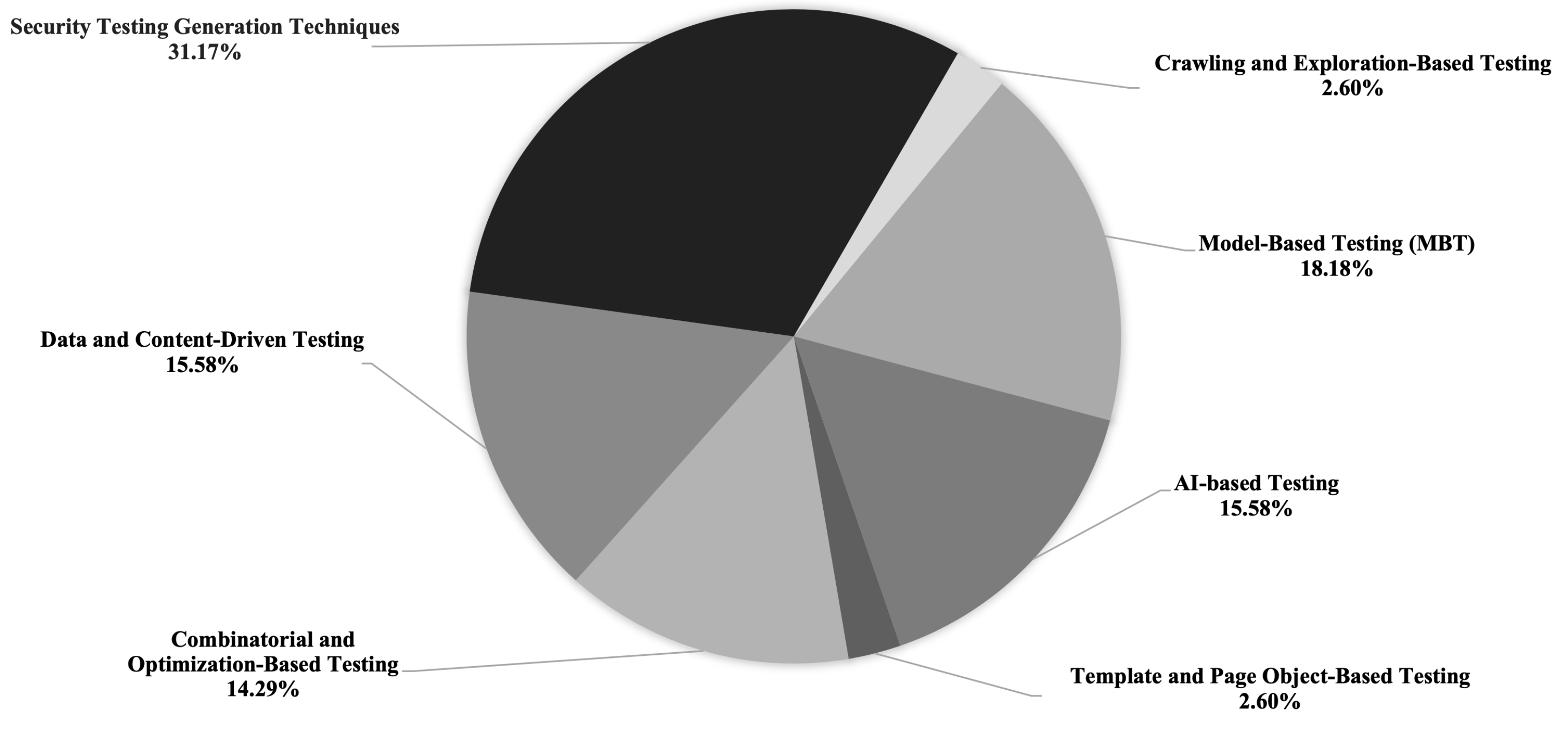}
        \caption{Proportions of generation methods in surveyed papers.}
    \label{Fig:test_case_generation_methods}
\end{figure}

Figure~\ref{Fig:test_case_generation_methods} presents the distribution of WAT test case generation methods identified in the surveyed literature. 
Security-oriented test case generation represents the largest category, accounting for 31.17\% of the papers. 
Model-based testing (MBT) ranks second at 18.18\% and derives test cases from models, such as UML diagrams and statecharts, that represent the structure or behavior of a system. 
Data- and content-driven testing and AI-based testing each account for 15.58\%
---
The former generates tests based on application data and content to evaluate input validation under diverse scenarios, whereas the latter uses AI techniques to support or automate test generation. 
Combinatorial and optimization-based testing accounts for 14.29\% and seeks to improve coverage by selecting effective combinations of input variables.
Template- and Page Object-based testing and crawling- and exploration-based testing are the least frequently reported categories, each accounting for 2.60\% of the papers.

\subsubsection{MBT}
Model-Based Testing (MBT) automatically generates test cases, enhancing both efficiency and coverage.
UML diagrams, statecharts, ER diagrams, and BPMN models are commonly used to represent the system and to generate the concrete test cases. 
\citet{panthi2017approach} and Suhag \& Bhatia~\cite{nabuco2014model} used UML models;
\citet{indumathi2016web} used ER diagrams and state transition models for database testing. 
\citet{akpinar2020web} showed how MBT could enhance coverage of complex interactions by integrating GraphWalker, Selenium, and JUnit. 
\citet{de2017test} used BPMN models to generate automated test scripts that covered functional test paths in Business Process Management (BPM) applications. 
\citet{wang2016method} proposed an approach based on UML statecharts for generating web link security test scenarios.
\citet{thummala2016using} used Petri nets to detect concurrency behaviors. 
\citet{milani2014leveraging} and \citet{waheed2020model} showed how model inference and script automation could enhance automation and test coverage.
\citet{muneer2025meta} used IFML constructs to organize checklist-based cross-browser compatibility test cases.
\citet{sathyakumar2024enhancing} used state machines to generate web UX component test paths.
\citet{wibowo2024extended} combined EFSM models with the Postman Problem algorithm to cover states and transitions in a web application workflow.

\subsubsection{Combinatorial and Optimization-Based Testing}
Combinatorial- and optimization-based techniques can generate efficient test cases, reducing the number of tests while improving coverage.
\citet{qi2017automated} introduced an automated approach based on combinatorial strategies:
This enhanced the coverage of dynamic pages.
\citet{bozic2015evaluation} reported on the effectiveness of IPOG and IPOG-F combinatorial testing algorithms for detecting XSS vulnerabilities.
\citet{wang2023parallel} and \citet{wang2019test} applied evolutionary algorithms and combinatorial testing methods to generate test cases covering complex paths in web applications:
Their work demonstrated the potential for genetic algorithms and simulated annealing in web testing. 
\citet{wang2016using} used combinatorial testing to construct navigation maps for dynamic web applications, enhancing the test coverage and effectiveness. 
\citet{guo2024multi} proposed a multi-objective generation approach that prioritizes important web states while maintaining test-case diversity under limited testing resources.
\citet{huang2024revealing} used response-time-guided genetic fuzzing to generate web API inputs that expose latency-inducing parameter combinations.

\subsubsection{AI-based Testing}
Testing methods can also use AI techniques to analyze and generate web application behaviors and inputs, automating the creation of effective test cases. 
\citet{chang2023reinforcement} proposed the WebQT tool, which leverages reinforcement learning to significantly improve testing efficiency and coverage through intelligent path selection.
\citet{pavanetto2020generation} used Generative Adversarial Networks (GANs) to generate highly realistic web navigation paths, effectively simulating real user operations. 
\citet{chaleshtari2023metamorphic} demonstrated the potential of AI techniques in security testing by using mutation testing to detect security vulnerabilities in web systems. 
\citet{shahbaz2015automatic} introduced a technique for generating test data using regular expressions and web searches, producing both valid and invalid WAT data. 
\citet{li2014symjs} developed the SymJS framework, which combines symbolic execution and dynamic feedback to improve the coverage and accuracy of JavaScript WAT.
\citet{leotta2024ai} evaluated ChatGPT and Copilot for translating Gherkin scenarios into Selenium-based E2E test scripts. 
\citet{abideen2025intent} proposed an approach based on LLMs for generating intent-driven Selenium E2E tests. 
\citet{wang2024leveraging} employed an LVLM to generate semantically meaningful inputs for web GUIs and determine subsequent interaction actions. 
\citet{nettur2024cypress} investigated the automated generation of Cypress test code from BDD scenarios.

\subsubsection{Data and Content-Driven Testing}
Data and content-driven testing methods aim to cover all possible input combinations, to comprehensively verify application functionality and security. 
\citet{hanna2019approach} proposed a method that analyzes client-side user input fields, effectively covering a range of input scenarios. 
\citet{hanna2018test} used semantically invalid input data to ensure that web applications reject inputs that do not meet semantic constraints, thereby enhancing security. 
\citet{shahbaz2015automatic} combined regular expressions with web searches to automatically generate test cases, verifying string validation logic in web applications. 
\citet{nagowah2017automated} developed the CRaTCP tool, which automatically extracts control combinations and generates test cases, achieving comprehensive WAT coverage. 
\citet{mariani2014link} used application data to automatically generate valid input data (that satisfy syntactic and semantic requirements) for testing complex web applications and ensuring coverage of input combinations.
\citet{wanigasekara2024comparing} compared adaptive genetic algorithms with LLM- and RAG-based techniques for automated test-data generation in web applications. 
\citet{aekaukkharachawakit2024generating} generated Robot Framework test scripts and associated test data from constraints obtained from front-end components and databases. 
\citet{quan2025tensorjsfuzz} extracted constraints from source code to generate valid inputs for web-based deep-learning frameworks.

\subsubsection{Crawling and Exploration-Based Testing}
Methods based on crawling and exploration are particularly well suited to testing complex navigation and state changes in dynamic web applications.
\citet{milani2014leveraging} combined automated crawling with manually written test cases to enhance coverage and explore unexplored paths. 
\citet{balera2022multiperspective} proposed a multi-perspective crawling approach using generative hyper-heuristic algorithms to capture diverse user behaviors and generate comprehensive test cases. 
\citet{yousaf2019automated} integrated model-driven techniques with crawling strategies to systematically navigate interfaces and improve test case quality.
\citet{de2015leveraging} leveraged task-driven methods to generate navigation paths aligned with functional workflows. 
\citet{nguyen2022generating} applied Q-learning to explore dynamic state spaces, effectively detecting vulnerabilities like XSS. 
\citet{pavanetto2020generation} utilized deep learning models that combined RNNs and GANs to simulate realistic navigation paths, thereby enhancing test coverage and interaction fidelity. 
Similarly, \citet{de2024automatic} proposed an automated approach for identifying distinct actionable elements across dynamic web states, thereby supporting systematic GUI exploration and test generation.

\subsubsection{Template and Page Object-Based Testing}
Robust and efficient test cases can be generated using templates or the Page Object Model.
\citet{yu2015incremental} used an incremental testing approach based on the Page Object Model that automatically generates page objects to represent web pages.
\citet{de2022morpheus} developed the Morpheus web-testing tool, which extracts UI component information from JSF and PrimeFaces frameworks to generate system-level functional test cases, effectively covering complex user interaction scenarios. 
\citet{leotta2015meta} employed metaheuristic techniques, including greedy and genetic algorithms, to automatically generate robust XPath locators, thereby reducing the maintenance overhead associated with DOM evolution.

\subsubsection{Security Testing Generation Techniques}
There are also techniques for generating test cases to detect security vulnerabilities in web applications.
Common attack types include XSS and SQLi.
\citet{akrout2014automated} proposed an automated black-box method for web vulnerability identification and attack scenario generation, thereby enhancing vulnerability detection and accuracy. 
\citet{awang2019method} used Cartesian product algorithms~\cite{agesen1995cartesian} to generate SQLi test data, effectively detecting vulnerabilities. 
\citet{wang2016method} used UML statecharts to generate web link security test scenarios, enhancing robustness in complex security environments.  
\citet{nguyen2022generating} used Q-learning~\cite{jang2019q} to detect security vulnerabilities and demonstrate the potential of intelligent algorithms in web security testing. 
\citet{wang2019test} used evolutionary testing, combining client-side and server-side testing scenarios to improve coverage and efficiency, and demonstrated the potential of combinatorial optimization algorithms in web security testing.
Prior studies have diversified security-oriented test generation through fuzzing, evolutionary search, request replay, reinforcement learning, and LLMs. 
\citet{yan2024automated} proposed property-aware fuzz-input generation for detecting data-binding vulnerabilities in Java web frameworks. 
For WAF evaluation, \citet{liang2023generative} employed transformer-based reinforcement learning to generate security-testing payloads, whereas \citet{yang2024llmsqli} used LLM agents to generate SQLi payloads for black-box web testing.
At the HTTP-request level, \citet{liu2025bacscan} constructed and replayed requests to detect broken access control, while 
\citet{chen2025racedb} generated and replayed candidate racing requests to expose race-condition vulnerabilities in database-backed web applications. 
\citet{lin2024spafuzz} combined fuzzing with genetic algorithms to generate SQLi and XSS test cases, and 
\citet{brandi2024sniping} generated rule-driven HTTP test cases targeting input-handling vulnerabilities. 
Finally, \citet{chen2024webfuzzauto} integrated reinforcement learning with LLM-generated strategies to guide web security fuzzing.

\section{ANSWER TO RQ3: Test Case Execution
\label{SEC:rq3}}
This section addresses RQ3 by examining how test cases are executed in the context of WAT. 
This section examines WAT execution methods, tools, platforms, environments, and related considerations.
Each aspect is analyzed in terms of its practical application and contribution to test-execution effectiveness.

\subsection{Environments}

The selection of testing environments is critical for WAT effectiveness, directly influencing the accuracy and scope of performance assessments. 
According to the literature, testing environments can be categorized into local~\cite{wang2023parallel}, remote~\cite{calzavara2021measuring}, and cloud-based~\cite{wei2019webhound}.

\begin{figure}[!t]
    \centering
        \includegraphics[width=0.6\textwidth]{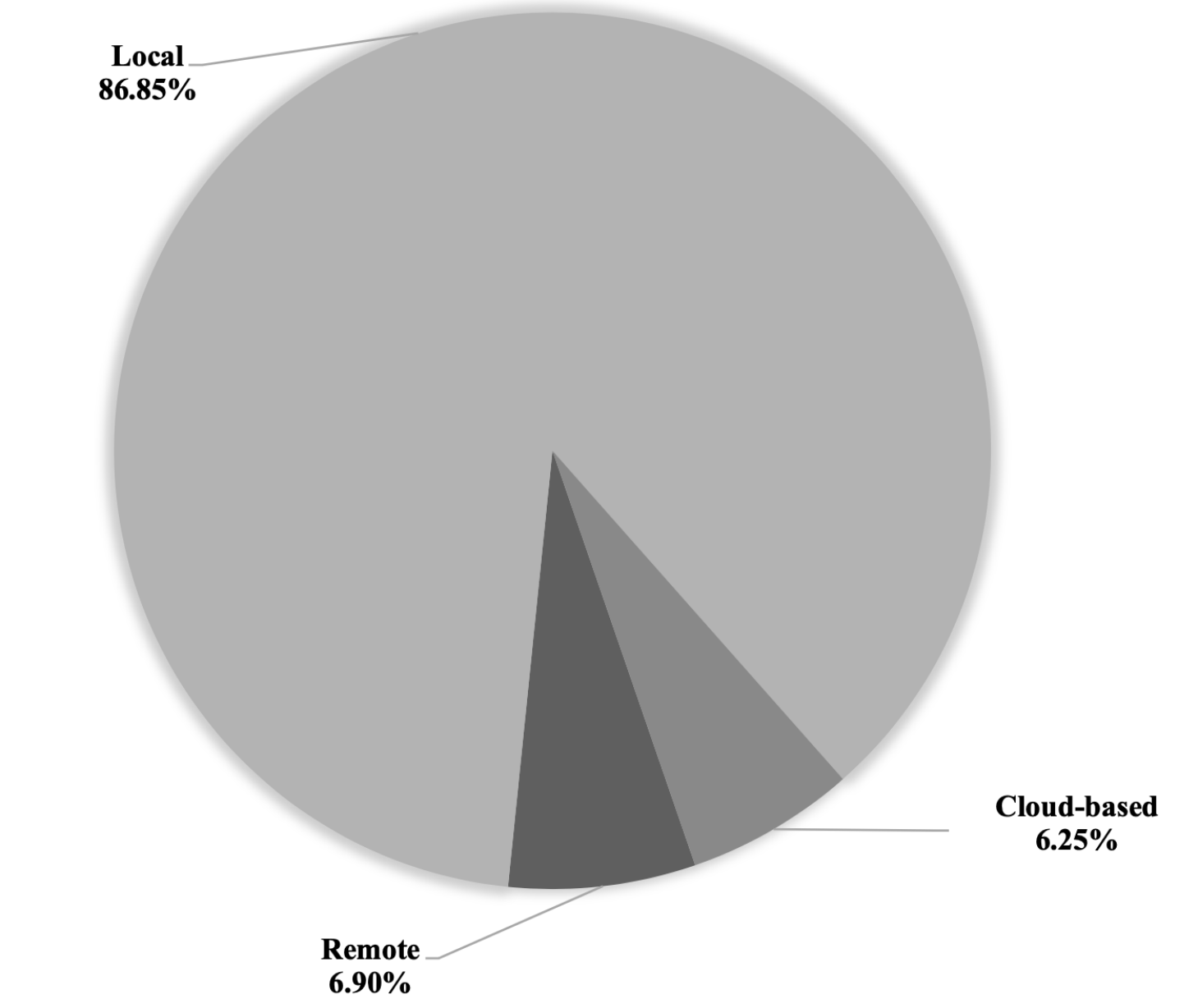}
    \caption{Proportions of execution environments in surveyed papers.}
    \label{Fig:test_case_execution_environments}
\end{figure}

Figure~\ref{Fig:test_case_execution_environments} shows that local environments were the most frequently used, accounting for 86.85\%. 
The control and immediate feedback potential of local environments make them particularly suitable for iterative testing in early development~\cite{everett2007software}. 
Remote environments (6.90\%) support testing under various network conditions:
This is essential for applications with geographically dispersed users.
Cloud-based environments (6.25\%) offer scalability and flexibility to meet complex testing demands and support continuous integration in large-scale applications. 

\subsubsection{Local Environments} 
The initial stages of development and testing often take place in local environments, which provide a controlled and isolated setting that facilitates rapid iteration and debugging.
In these environments, because web applications are deployed on the same machine where tests are executed, the speed of iteration and bug fixing can be increased.
Tools like Apache JMeter are increasingly adopted for testing web applications in local environments~\cite{tiwari2023analytical}. 
JMeter excels at performance and functional testing and provides a reliable framework for validating isolated components during unit and integration testing, especially in controlled local environments.
This controlled setting also supports testing activities beyond conventional performance and functional evaluation:
Local browser-based execution has been applied to accessibility mutation analysis and responsive web accessibility assessment~\cite{tafreshipour2024ma11y,chiou2024automatically}.
Moreover, LLM-assisted accessibility repair can integrate browser-based inspection components, such as Axe-Playwright, with model-driven corrections to rendered HTML~\cite{fathallah2025accessguru}.

A challenge for local environments, however, is that they cannot easily simulate real-world conditions such as network latency, distributed-system behavior, and large-scale load and performance conditions~\cite{prisadi2023implementation}.
To address these limitations, tools such as Docker have been widely used to create containerized environments that maintain consistency across development and production stages~\cite{halin2019test}.
Docker ensures that testing environments are uniform, thus reducing discrepancies among development, testing, and production setups~\cite{marquardt2021deja}.
Security-oriented WAT studies further demonstrate the importance of local environments:
Such environments allow researchers to configure vulnerable applications, inspect source code and database states, replay requests, and validate detected vulnerabilities under controlled and reproducible conditions.
Representative examples include RecurScan for detecting recurring PHP vulnerabilities~\cite{shi2024recurscan}, BACScan for black-box testing of broken access control~\cite{liu2025bacscan}, and studies addressing missing-owner-check, request-race, taint-style, and Java data-binding vulnerabilities~\cite{liu2025mocguard,chen2025racedb,liu2025detecting,yan2024automated}.

\subsubsection{Remote Environments}
Remote environments extend local capabilities by enabling test cases to run across multiple physical or virtual machines.
Tools like Selenium Grid~\cite{balera2022multiperspective} play an important role in remote environments, enabling cross-browser and cross-platform testing~\cite{debroy2018automating}. 
Selenium Grid helps ensure that web applications function consistently across different environments and configurations~\cite{jana2019appmine}.
Remote environments are particularly suitable for testing targets that cannot be faithfully reproduced locally, including live websites, public web services, and ecosystem-scale deployments.
For example, large-scale measurements of web cache poisoning require observations of real-world Internet infrastructure~\cite{liang2024internet}, whereas accessibility studies of externally hosted institutional websites rely on archived or deployed pages whose content may evolve independently of the testing environment~\cite{nso2025impact}.

Despite their advantages, remote environments also face challenges, including network latency and the need to synchronize test execution across different systems~\cite{zhou2014ltf}. 
Docker, in conjunction with Selenium Grid, helps alleviate these issues by ensuring consistent test environments across remote nodes~\cite{halin2019test}.
Remote execution also introduces challenges to reproducibility:
Tested pages, third-party scripts, service policies, authentication states, and network conditions may change during data collection.
Consequently, remote WAT typically requires comprehensive logging, well-defined retry policies, timestamped observations, and a clear distinction between failures attributable to the application under test and those caused by its surrounding service environment.
These concerns are particularly evident in hybrid settings, where local test scripts or browser controllers interact with remote pages, third-party trackers, multilingual web content, or externally hosted model services~\cite{zeng2025measuring,bhuiyan2025not,le2025embedrank}.

\subsubsection{Cloud-Based Environments} 
Cloud-based environments offer the most flexible and scalable solution for large-scale WAT.
Platforms such as \textit{Amazon Web Services} (AWS), \textit{Google Cloud Platform} (GCP), and Microsoft Azure offer dynamic resource allocation, allowing tests to be executed on demand across multiple geographical locations~\cite{ludinard2018invariant,su2021elastic}.
A cloud environment can simulate real-world scenarios (including high traffic and various network conditions), and can support global users~\cite{arcuri2023emb}.
AWS Lambda, for example, is an important tool in cloud environments, providing a serverless architecture that enables tests to run without manual infrastructure management~\cite{mukherjee2014performance}.
It can also dynamically expand resources based on test execution requirements, ensuring efficient test runs.
Although cloud platforms are less commonly adopted as the primary execution environment, they remain important in several specialized scenarios.
Examples include CI/CD-based testing and optimization~\cite{thokala2025testing}, test-case optimization simulations conducted in cloud notebook environments~\cite{manojkumar2024test}, and the cloud deployment of phishing-detection web applications~\cite{ghadiyaram2025phish}.

\subsection{Execution Methods}

Test case execution plays a critical role in verifying web application functionality, security, and performance. 
Various methods can be used to execute test cases, each with advantages depending on the application's nature and complexity. 
This section provides a detailed exploration of the three primary approaches to test-case execution: 
manual, automated, and hybrid. 
Each approach is analyzed for its practical application, highlighting each approach's contribution to the overall effectiveness of WAT execution.

\begin{figure}[!t]
    \centering
        \includegraphics[width=0.6\textwidth]{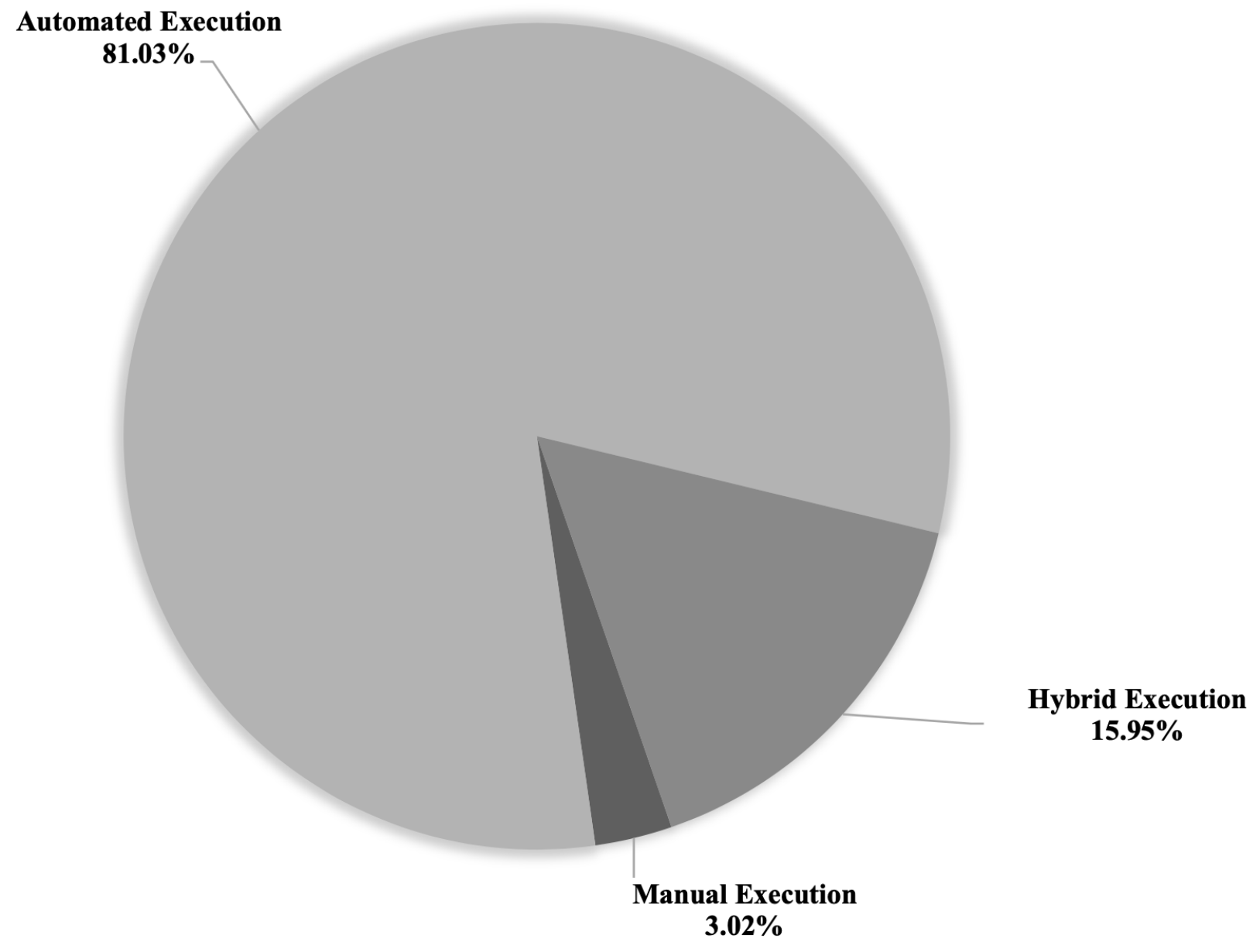}
    \caption{Proportions of execution methods in surveyed papers.}   
    \label{Fig:test_case_execution_methods}
\end{figure}

Figure~\ref{Fig:test_case_execution_methods} presents an overview of the different approaches to test-case execution in the surveyed literature.
Automated execution dominates, appearing in 81.03\% of the surveyed papers.
This is as expected, given its efficiency and scalability for repetitive tasks and regression testing.
Automated execution can support tasks requiring consistency across environments.
Manual execution, although present in only 3.02\% of the surveyed papers, remains essential for tasks such as exploratory testing and scenarios requiring human intuition.
Hybrid execution, accounting for 15.95\%, combines both approaches, optimizing test coverage and flexibility by automating routine tasks while leveraging manual testing for subjective analysis.
This can be particularly valuable in complex testing scenarios where both automation efficiency and human judgment are required.

\subsubsection{Manual Execution}
Manual execution involves human testers directly interacting with web applications to verify their behavior. 
This can be particularly effective for exploratory testing, as testers need the flexibility to respond dynamically to application behavior and detect unexpected issues.
Exploratory testing can help identify user experience issues or unexpected errors that are difficult to predict when designing automated scripts~\cite{fan2023comprehensive}.

Manual testing is also often used to evaluate new features or complex user interfaces, which can require human judgment to assess usability, aesthetics, and overall user satisfaction~\cite{leotta2020family}. 
However, there can be significant limitations to manual testing, including its time-consuming, labor-intensive nature and susceptibility to human error.
This is especially true when dealing with large applications that require executing hundreds of test cases~\cite{garg2015framework}.
Therefore, although manual testing remains essential for specific tasks, it may be most suitable for early development stages or areas that require subjective validation~\cite{gupta2014keyword}.

The importance of manual execution is particularly evident when test outcomes depend on direct user interaction or domain expertise rather than predefined pass/fail assertions.
For example, manual usability evaluations and validation sessions have been conducted for health decision-support tools, medical-scoring applications, and web-based training systems to assess their suitability for clinical or educational workflows~\cite{moseley2025usability,lang2025orthodontic,nuruzzaman2025preliminary}.
Similarly, manual accessibility evaluation can uncover barriers associated with motor effort, posture, repetitive movements, and stress that rule-based checkers may overlook~\cite{gallagher2024motor}.

\subsubsection{Automated Execution}
Automated execution provides a more scalable and efficient approach, particularly when test cases must be repeatedly executed across different environments.
Typically, scripts and tools are used to execute test cases, capture results, and compare them with predefined expected outcomes~\cite{leotta2020family}.
This is often done without the need for further human intervention.
Selenium, for example, is widely used to automate functional testing across various browsers, ensuring consistent behavior and compatibility~\cite{appelt2018machine}.

The scope of automated execution has also expanded through the integration of LLM-assisted and model-based techniques.
These techniques can generate or refine executable web tests, which are subsequently run using tools and frameworks such as Selenium and JUnit~\cite{leotta2024ai,wang2024leveraging}.
Beyond functional testing, automated execution has been applied to accessibility assessment~\cite{chiou2024automatically,tafreshipour2024ma11y,bhuiyan2025not,fathallah2025accessguru}, as well as security, privacy, and performance analysis~\cite{xie2024arcanum,chen2025racedb,di2025performance}.

Automated execution can significantly improve efficiency by reducing human involvement and allowing tests to run continuously in different environments.
However, automated execution can require considerable upfront investment in scripting and tool configuration~\cite{ammann2017introduction}.
Furthermore, it may not be suitable for subjective analysis, such as in visual interface testing~\cite{garg2015framework,lounis2014new}.

\subsubsection{Hybrid Execution}
Hybrid execution combines manual and automated techniques, drawing on the strengths of each.
In hybrid models, manual testing is typically used for exploratory tasks or validation of new features, while automated scripts handle repetitive tasks and regression testing~\cite{pelivani2021comparative}.
This approach offers flexibility by allowing testers to manually address complex, subjective scenarios while leveraging automation for repetitive, time-consuming processes~\cite{guan2021comparative,garg2015framework}.

It has been reported that combining UML activity diagrams with manual and automated test cases could improve overall test coverage and shorten execution time~\cite{zhao2015new}. 
Similarly, a hybrid approach can be adopted in security testing, in which potential vulnerabilities (such as SQLi points) can be detected automatically, and manual testers can then validate the findings~\cite{singh2020automated}.
The combination of the thoroughness of manual testing and the efficiency of automation 
often makes hybrid execution the most practical solution
for complex and large-scale web applications~\cite{gupta2014keyword}.

Hybrid execution should be distinguished from fully automated testing, in which human involvement is limited to reviewing results after execution.
In security testing, for example, automated techniques can first identify potential vulnerabilities or suspicious behaviors, which are subsequently assessed by experts~\cite{shi2024recurscan,liu2025mocguard,liang2024internet}.
Human judgment may also be incorporated directly into automated workflows to calibrate test oracles, refine decision thresholds, or guide LLM-assisted testing strategies~\cite{lopez2025turning,olianas2025leveraging}.
This pattern combines the scalability and repeatability of automation with the contextual judgment required for complex testing decisions.

\section{ANSWER TO RQ4: Evaluations and Metrics
\label{SEC:rq4}}

This section addresses RQ4 by exploring WAT metrics and evaluation methods. 
We first discuss the primary effectiveness metrics, such as test case quality, coverage, failure-detection capability, and vulnerability detection. 
This is followed by an analysis of efficiency metrics, such as the time required for generation, execution, and error detection. 
Finally, empirical studies are reviewed to explain how these metrics have been applied to evaluate and improve the effectiveness and efficiency of WAT.

\subsection{Test Effectiveness}

Test effectiveness evaluates the ability of WAT methodologies to identify faults, vulnerabilities, and failures. 
This involves using specific metrics, including test-case quality, code coverage, failure-detection capability, and vulnerability-detection effectiveness. 
These metrics provide insights into the thoroughness and accuracy of WAT processes.

\begin{figure}[!t]
    \centering
        \includegraphics[width=0.7\textwidth]{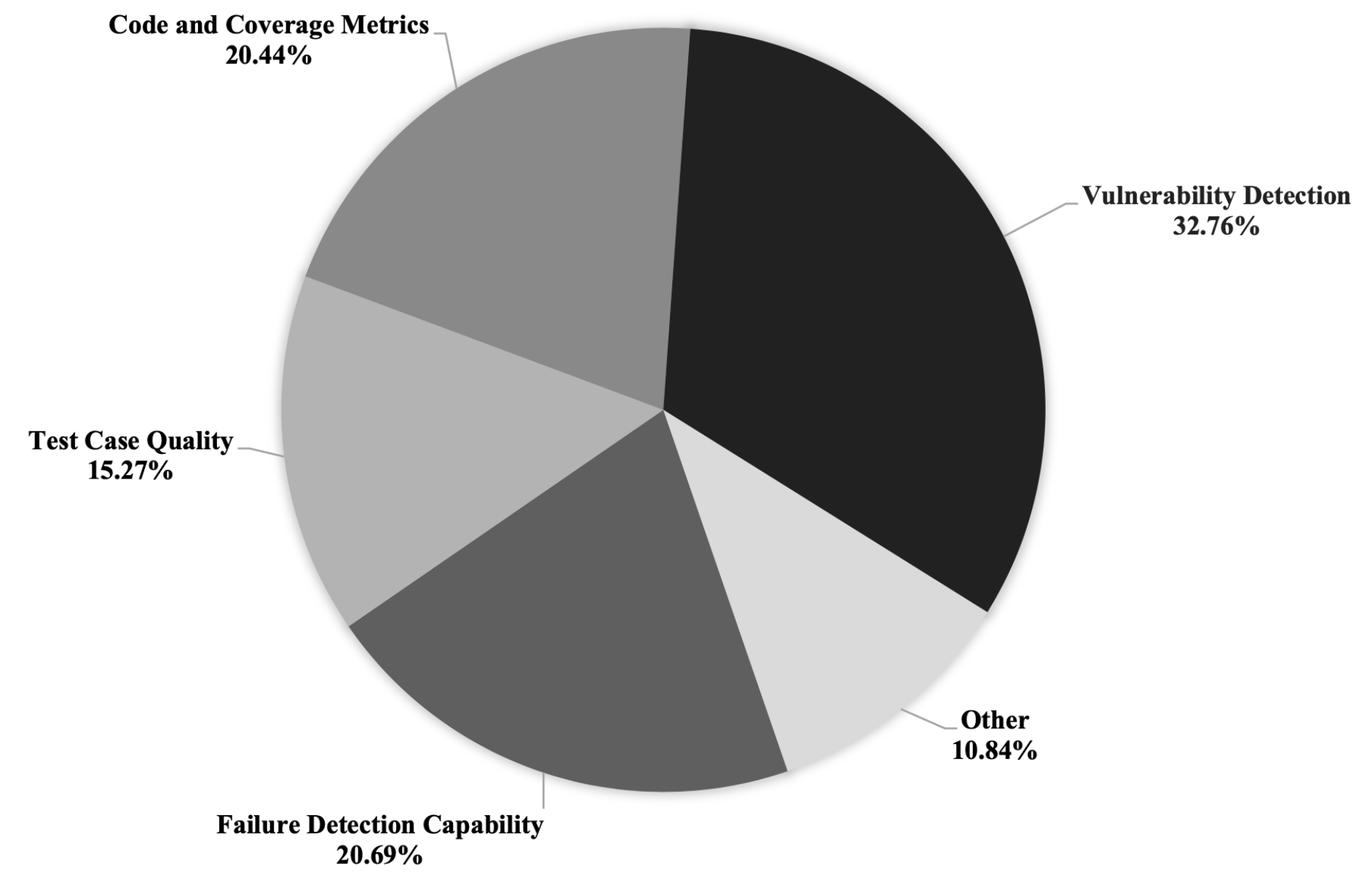}
    \caption{Proportions of test effectiveness metrics in surveyed papers.}
    \label{Fig:test_effectiveness_rate}
\end{figure}

Figure~\ref{Fig:test_effectiveness_rate} presents an overview of the different test effectiveness metrics used in the surveyed literature.
Vulnerability detection is the most prevalent category, representing 32.76\% of all identified effectiveness metrics.
Failure-detection capability and code and coverage metrics account for 20.69\% and 20.44\%, respectively, with the latter reflecting the breadth of test execution.
Test-case quality represents a further 15.27\%, while the remaining 10.84\% comprises other types of metrics.
For example, \citet{saharan2018analytical} and \citet{tiwari2023analytical} refer to performance-based metrics such as response time, error rate, and throughput.

\subsubsection{Test-Case Quality}
The quality of test cases directly affects the ability to detect defects in web applications.
Dynamic testing approaches, such as Q-learning-based frameworks, have demonstrated substantial improvements in test-case adaptability. 
These methods optimize element interaction paths, increase branch and element coverage, and effectively identify critical defects in complex application environments~\cite{fan2023comprehensive}. 
These tools provide a scalable solution for testing modern web applications by enhancing exploratory capabilities.

Test-case quality can also be evaluated in terms of the executability of generated inputs and the stability of GUI interactions.
TensorJSFuzz, for example, assesses input validity according to whether the generated inputs can be executed successfully~\cite{quan2025tensorjsfuzz}.
Similarly, research on waiting strategies for Selenium tests uses completion rates and interaction stability as indicators of test reliability~\cite{zhang2024towards}.
Prioritizing test cases based on risk and frequency of modification has enhanced the efficiency of defect detection. 
These strategies uncover critical defects earlier by focusing on high-risk and frequently modified components, thereby streamlining testing processes and improving efficiency and effectiveness~\cite{ali2015case}.

Furthermore, fast evaluation techniques that leverage clustering-based sampling can improve test-case efficiency in large-scale applications. 
These methods group similar application paths and ensure representative coverage, which significantly reduces redundant testing efforts while maintaining reliability~\cite{zhang2015sampling}.

\subsubsection{Code Coverage Metrics}
Code coverage metrics
---
including line, branch, and function coverage
---
can help show how thoroughly an application’s codebase is exercised during testing~\cite{nguyen2019exploring,wang2016using}.
Coverage analysis tools such as Selenium enable testers to capture interactions across multiple layers of a web application, from the front-end UI to the back-end logic~\cite{paul2018approach}.
Coverage metrics have also been extended beyond source code to include GUI states and actions, thereby measuring the range of application states and user interactions explored during automated web testing~\cite{fan2024can,wang2024leveraging}.
In accessibility testing, coverage has similarly been used to quantify the breadth of accessibility-violation categories detected by a testing approach~\cite{fathallah2025accessguru}.

However, high coverage alone does not guarantee comprehensive fault detection. 
Research has shown that combining coverage metrics with fault injection methods enhances the ability to uncover defects~\cite{chaleshtari2023metamorphic}.
Injecting controlled faults into the system and observing whether or not the test cases can detect them provides a more rigorous assessment of a test suite’s fault-detection capabilities~\cite{gupta2014keyword}.
Mutation testing complements code coverage metrics by introducing controlled changes (i.e., ``mutants'') to evaluate the fault-detection capabilities of test cases~\cite{habibi2015event, leotta2024mutta}.
While coverage metrics measure the breadth of testing, mutation testing ensures that covered paths are rigorously tested, enhancing the reliability of the testing process.

\subsubsection{Failure-Detection Capability}
Failure-detection capability refers to the accurate identification and localization of faults during testing. 
Techniques such as dynamic trace-based analysis compare the execution traces of different versions of the application to identify discrepancies~\cite{touseef2019analysis}.
This approach has proven effective for failure detection, particularly when combined with automated tools that analyze execution logs to identify potential issues, such as security vulnerabilities, in large-scale applications.
This can help testers to pinpoint the specific lines of code or configuration changes responsible for faults, significantly reducing the time required for debugging and resolution~\cite{walsh2015automatic,halin2019test}.

Failure-detection capability can also be evaluated through controlled fault injection and comparisons across multiple executions.
For example, mutation-based accessibility testing examines whether deliberately injected accessibility faults can be successfully detected~\cite{tafreshipour2024ma11y}.
Execution-based approaches can similarly detect and validate failures by comparing system behavior or observations obtained under different execution conditions~\cite{chen2025racedb,liang2024internet}.

Automated diagnostic tools further improve this process by analyzing execution logs and error traces to identify the exact points of failure~\cite{sunman2022automated}.
These tools are particularly useful for large-scale web applications, where manual fault localization can be time-consuming and error-prone~\cite{fan2023comprehensive}.

\subsubsection{Vulnerability Detection}
Vulnerability detection involves identifying potential security issues, such as SQLi and XSS, that could be exploited by attackers~\cite{awang2019method,mokbal2019mlpxss}.
Automated vulnerability detection tools simulate real-world attack scenarios to test the robustness of the application’s security defenses~\cite{nagpure2017vulnerability}.
Vulnerability-detection effectiveness can be quantified using metrics such as precision, recall, and false-positive rates, as well as an approach's ability to uncover previously unknown vulnerabilities.
Existing studies have employed such measures to evaluate vulnerability-detection techniques across a range of web application and security contexts~\cite{liu2025bacscan,liu2025mocguard,liu2025detecting,shi2024recurscan,xie2024arcanum}.

Machine-learning-based techniques have further strengthened automated vulnerability detection.
These systems analyze traffic and behavioral patterns to identify anomalies that may indicate vulnerabilities~\cite{jana2019appmine}.
As machine learning models learn from new data, they continuously improve their ability to detect both known and unknown threats, providing a dynamic defense against emerging security risks~\cite{calzavara2020machine}.

\subsection{Test Efficiency}

\begin{figure}[!t]
    \centering
        \includegraphics[width=0.7\textwidth]{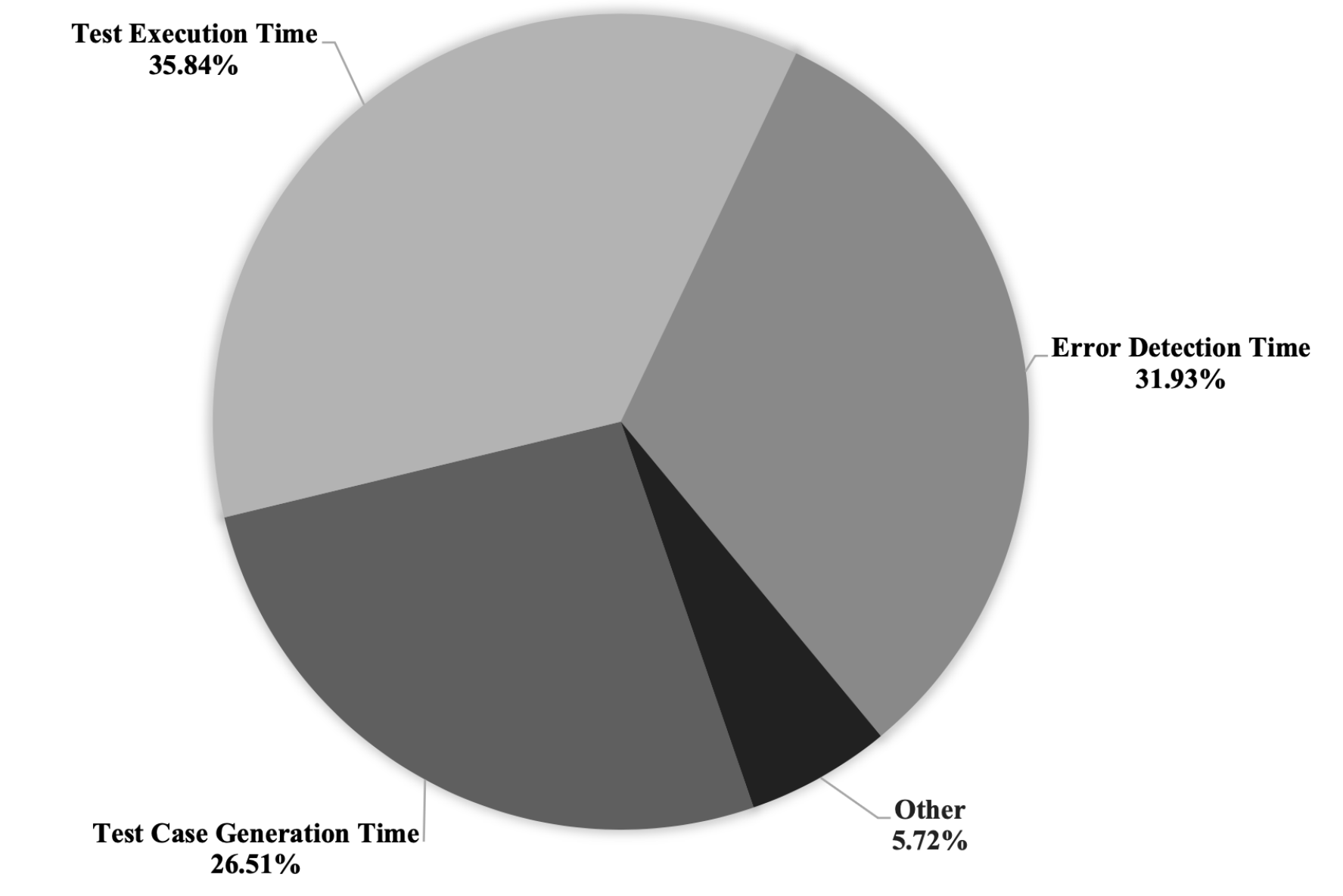}
    \caption{Proportions of test efficiency metrics in survey papers.}
    \label{Fig:test_efficiency_rate}
\end{figure}

Test efficiency is a core WAT metric that directly affects the cost-effectiveness of the testing process.
Efficiency metrics assess the resources required for comprehensive test coverage and fast fault detection.
Some of the main metrics include the time taken for test case generation~\cite{gupta2014keyword} and execution~\cite{pelivani2021comparative}, and error detection~\cite{fan2023comprehensive}.
Figure~\ref{Fig:test_efficiency_rate} summarizes the test-efficiency metrics reported in the surveyed literature.
Test-execution time is the most frequently measured metric, representing 35.84\% of all identified efficiency metrics.
Error-detection time accounts for a further 31.93\%, followed by test-case generation time at 26.51\%.
The remaining 5.72\% comprises other efficiency measures.
For example, \citet{sung2016static} employed static analysis to eliminate redundant tests, thereby improving testing speed and overall efficiency.

\subsubsection{Test Case Generation Time}
The test-case generation time measures how efficiently test cases are created and is a major factor in overall testing efficiency.
Several techniques have been developed to streamline this process. 
The Page Object Pattern (POP) has also reduced test-case generation time as the test suite grows by reusing defined objects. 
This approach improves efficiency by reusing the same components across multiple test cases~\cite{leotta2020family}. 
Studies involving WebQT, which employs machine learning algorithms, have shown that test-case generation can be significantly accelerated through optimized exploration and test-sequence generation~\cite{gupta2014keyword}. 
Test-case generation efficiency can also be assessed in terms of the effort required to develop executable test scripts, particularly when AI-assisted generation techniques are employed~\cite{leotta2024ai}.
More effective constraint extraction can further improve generation efficiency by increasing the proportion of valid test inputs and reducing effort spent on unsuccessful generation attempts~\cite{quan2025tensorjsfuzz}.
Tools like CRaTCP can automatically generate and execute all possible test case combinations to further reduce manual effort~\cite{nagowah2017automated}.

\subsubsection{Test-Execution Time}
The test execution time is the time required to execute the generated test cases. 
Automated testing frameworks like Selenium and Katalon Studio can dramatically reduce execution time (compared to manual testing)~\cite{pelivani2021comparative}. 
Research has confirmed their effectiveness in optimizing test execution times, particularly in large-scale web applications~\cite{pelivani2021comparative}. 
Genetic algorithms, for example, have been used to reduce execution time by optimizing test paths, thereby enabling earlier fault detection~\cite{stallenberg2019jcomix}. 
Concurrent execution techniques, where multiple test cases are run in parallel, 
can further reduce execution times~\cite{garg2015framework}. 
Execution time is often evaluated alongside computational cost and throughput, particularly in large-scale web-testing systems~\cite{wu2025autopt,nie2025wascope}.
Replay-based techniques can further improve scalability by reducing the need for repeated interactions with live web environments~\cite{xie2024arcanum}.
Prioritization techniques that execute high-risk test cases first can lead to more efficient use of resources by focusing on the most critical components.

\subsubsection{Error-Detection Time}
The error-detection time reflects how quickly errors or vulnerabilities are identified during testing.
Hybrid methods that integrate static and dynamic analysis improve detection efficiency by focusing on critical code paths and minimizing redundant analysis~\cite{zhao2015new}. 
Techniques that leverage anomaly detection through multi-model approaches accelerate fault detection by identifying deviations in real time and ensuring low false-positive rates~\cite{zhang2017anomaly}.
By minimizing manual intervention, automated test suites enable real-time fault identification and significantly accelerate fault detection~\cite {paul2018approach}. 
In static analysis and fuzz testing, detection time is closely associated with exploration depth and the ability to generate valid inputs~\cite{meng2025slvhound,quan2025tensorjsfuzz}.
In fully automated penetration testing, detection latency is also related to task completion and the overall operational cost of the testing process~\cite{wu2025autopt}.
Moreover, combining diverse static analysis tools has significantly improved detection speed and accuracy, particularly in resource-constrained or time-sensitive scenarios~\cite{nunes2019empirical}.

\subsection{Empirical Studies}
This section examines the use of empirical studies in WAT, with a focus on effectiveness and efficiency metrics. 
It explains how these studies can enhance testing performance, optimize testing processes, and offer insights for further improvements in testing tools.

\subsubsection{Application of Effectiveness and Efficiency Metrics in WAT}
Coverage has been widely used to measure testing effectiveness in empirical WAT studies~\cite{garousi2021model,mattiello2022model}. 
MBT can generate high-quality test cases, with good code coverage and fault-detection capabilities~\cite{nabuco2014model,nguyen2019exploring}.
Compared with manual testing, MBT provides a more systematic approach to covering application requirements, especially for complex systems, and achieves greater testing efficiency~\cite{wang2019test}. 
MBT is particularly effective at handling the complex interactions between multiple modules or interfaces within large-scale web applications.
Empirical studies have shown that MBT reduces test redundancy and optimizes test case generation~\cite{hanna2018test}.

An automated tool such as Model-Based Testing Leveraged for Web Tests (MoLeWe)~\cite{mattiello2022model} has demonstrated improved testing efficiency~\cite{li2014modeling}:
MoLeWe can generate high-quality test cases, increasing test coverage while reducing execution time~\cite{garousi2021model}. 
It can significantly reduce manual testing time, especially in applications with frequent version updates.
MoLeWe has also demonstrated flexibility in resource allocation, optimizing system resource utilization in multi-threaded testing.

Fault Detection Density (FDD) and Fault Detection Effectiveness (FDE) are core effectiveness metrics for evaluating the fault-detection capabilities of testing tools~\cite{habibi2015event}. 
The MUTANDIS mutation-testing tool can generate non-equivalent mutants, significantly improving the fault-detection rate in JavaScript applications~\cite{mirshokraie2014guided}. 
Mutation testing can cover a wide range of fault scenarios using diverse mutants:
This can improve the comprehensiveness of the test cases. 
MUTANDIS has also been shown to enhance fault-detection efficiency while reducing resource waste caused by redundant testing paths.

Efficiency metrics are not only used for coverage and fault-detection capabilities, but also for execution time and resource utilization~\cite{putri2017performance,ali2015performance}.
Empirical studies have shown that tools such as JMeter can optimize performance testing using distributed architectures~\cite{saharan2018analytical,kiran2015experiences}.
By parallelizing test case execution, JMeter can significantly reduce execution time while ensuring system stability under high-load conditions.
This architecture has been shown to perform well in real-world scenarios, especially in large-scale multi-user concurrent environments~\cite{guarnieri2017test}.

Empirical evaluations have also accounted for testing effort and resource constraints when assessing testing efficiency.
Normalizing test-script development effort by test size enables more meaningful comparisons among scripts of varying lengths and levels of complexity~\cite{leotta2024ai}.
In resource-constrained environments, multi-objective test generation can further reduce the number of executed test cases while preserving coverage of important application elements~\cite{guo2024multi}.

\subsubsection{Improvements in Testing Performance through Empirical Studies}
Empirical studies have confirmed the effectiveness of reinforcement learning techniques in the prioritization of test cases~\cite{chang2023reinforcement}:
High-risk areas can be prioritized, thereby reducing overall testing time and improving fault-detection efficiency.

Automated tools can also improve testing efficiency.
MoLeWe can automatically generate test cases and scripts, minimizing manual intervention and significantly enhancing automation~\cite{mattiello2022model}:
This not only reduces the error rate associated with manual operations but also enables rapid adaptation of test suites to accommodate new features or updates.

Empirical findings further indicate that testing performance can be improved by balancing test quality against execution efficiency.
Increasing the proportion of valid generated inputs can accelerate fault detection, whereas eliminating redundant test cases can reduce execution effort without sacrificing coverage of important application elements~\cite{quan2025tensorjsfuzz,guo2024multi}.
For fully automated penetration testing, improvements in task-completion capability should also be evaluated alongside runtime and operational cost to identify and mitigate end-to-end testing bottlenecks~\cite{wu2025autopt}.

\section{ANSWER TO RQ5: Available Tools
\label{SEC:rq5}}

This section addresses RQ5 by identifying and cataloging the WAT tools currently available.
This should provide an overview of the technological resources available to support this field. 
We first list the tools and then classify them by type and features. 
The section concludes with some details on tool accessibility, including links for further exploration.

\subsection{Testing Tools}
Various WAT tools are available to streamline and automate testing, enhancing efficiency, scalability, and consistency across diverse environments. 
These tools can support the automation of various types of testing (including functional, regression, security, and performance testing), helping to examine the reliability and robustness of web applications.

\begin{figure}[!b]
    \centering
        \includegraphics[width=0.7\textwidth]{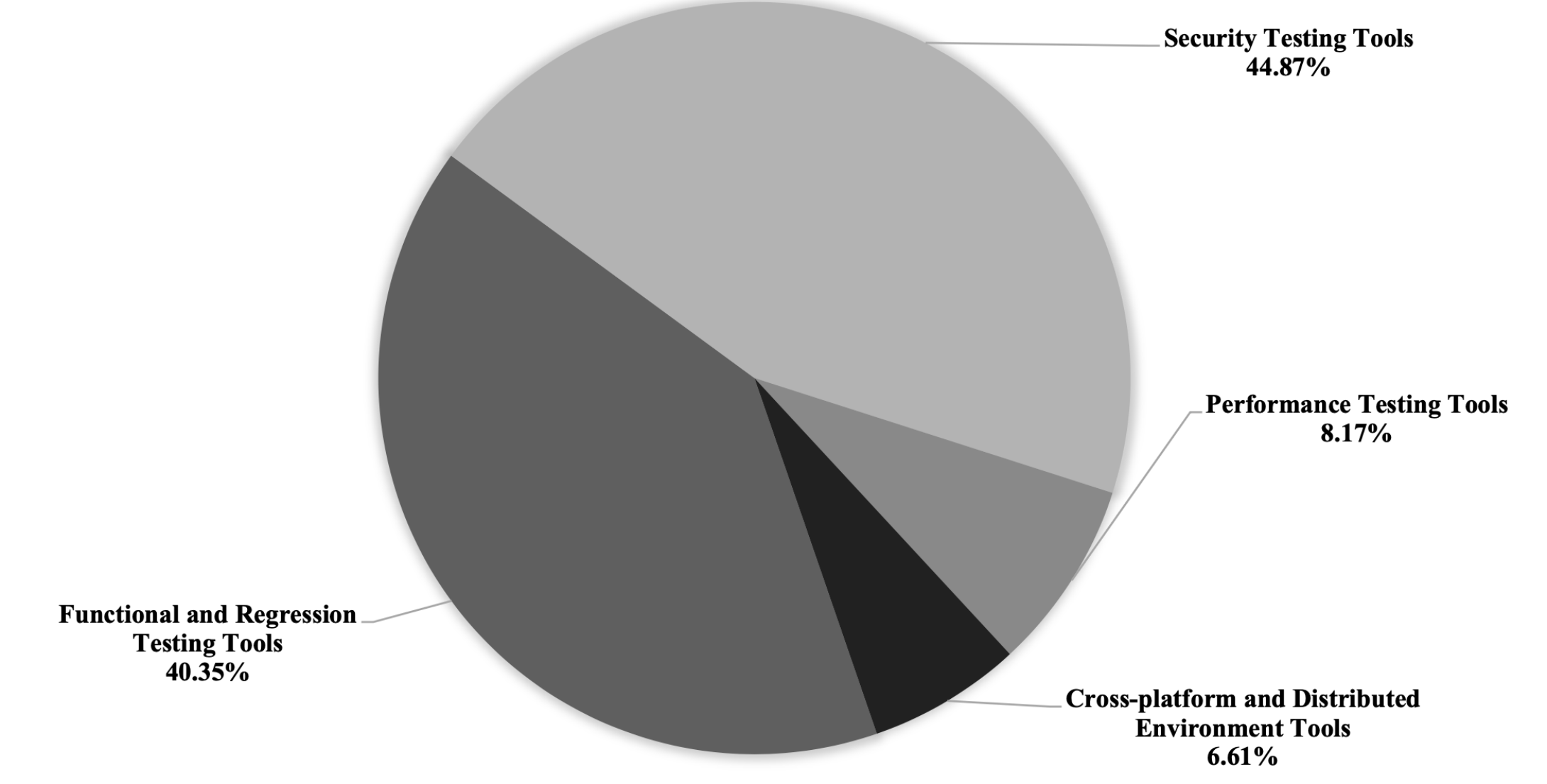}
    \caption{Proportions of testing tools in surveyed papers.}
    \label{Fig:testing_tools_rate}
\end{figure}

Figure~\ref{Fig:testing_tools_rate} summarizes the distribution of testing-tool categories identified in the surveyed literature.
Security testing tools constitute the largest category, accounting for 44.87\% and reflecting the substantial emphasis placed on mitigating cybersecurity risks.
Functional and regression testing tools represent a further 40.35\%, highlighting their importance in validating application behavior and maintaining system stability following software changes.
Performance testing tools account for 8.17\%, while cross-platform and distributed-environment tools account for 6.61\%, reflecting their more specialized roles in performance evaluation and consistent execution across platforms.
Overall, security and functional testing receive the greatest attention, whereas performance and cross-platform tools are employed more selectively.

\subsubsection{Functional and Regression Testing Tools}
Functional and regression testing tools, like Selenium~\cite{pelivani2021comparative} and Katalon Studio~\cite{pelivani2021comparative}, have been widely used for cross-browser and cross-platform testing:
By automating test execution across various browsers and operating systems, they help to verify that web applications function correctly after updates and code changes 
Browser automation frameworks have also been applied to test generation, usability evaluation, and web GUI testing~\cite{abideen2025intent,lu2025uxagent,zhang2024towards}.
Tools built on Playwright and Puppeteer additionally support automated accessibility evaluation by identifying accessibility issues in web applications~\cite{fathallah2025accessguru,tafreshipour2024ma11y,chiou2024automatically}.

\subsubsection{Security Testing Tools}
Tools like Burp Suite~\cite{nagpure2017vulnerability} and OWASP ZAP~\cite{amankwah2020automated} can help to identify security vulnerabilities, including SQLi, XSS, and other common web application threats. 
These tools are essential for protecting web applications from attacks and ensuring compliance with security standards.
Automated security-testing approaches have also been developed to detect different classes of vulnerabilities across a range of web applications~\cite{liu2025bacscan,chen2025racedb,shi2024recurscan}.
Other approaches integrate automated crawling, fuzzing, and LLM-based techniques to support penetration testing and vulnerability discovery~\cite{wu2025autopt,chen2024webfuzzauto,yang2024llmsqli}.

\subsubsection{Performance Testing Tools}
JMeter~\cite{fan2023comprehensive} and LoadRunner~\cite{khan2016smoke} have been used to simulate high-traffic loads, helping to evaluate how web applications perform under stress. 
They assess critical performance metrics, including response times and throughput, and evaluate whether the application remains stable under heavy load conditions.
The k6 tool has also been employed to automate load, spike, and soak testing by generating different traffic workloads~\cite{pinastawa2024web}.

\subsubsection{Cross-platform and Distributed Environment Tools}
Tools such as Docker~\cite{marquardt2021deja} and Selenium Grid~\cite{kiranagi2017feature} enable test cases to be executed in distributed environments and containerized platforms. 
They help to ensure consistency in execution and provide scalability for larger systems, supporting the efficient management of complex testing scenarios.
Docker-based containers and Kali Linux-based environments have also been used to support automated testing.
Meanwhile, CI/CD platforms such as AWS CodePipeline, Azure Pipelines, Jenkins, and GitLab CI/CD facilitate test execution across different deployment environments~\cite{wu2025autopt,thokala2025testing}.

\subsection{Tool Accessibility}
The ease of access to WAT tools is a crucial factor in their adoption and eventual effectiveness in the software development and testing community. 
Table~\ref{TAB:wat_tools} 
lists some popular WAT tools\footnote{To measure the popularity of each WAT tool, we collected the number of times it was mentioned in the selected 464 primary studies.}, categorized by type, core functionalities, and access links:
This structured presentation facilitates the selection of tools based on specific testing requirements, ranging from performance evaluation and security vulnerability detection to automated UI testing and code quality analysis. 
Due to page limits, we provide a comprehensive summary of nearly all relevant tools in the online supplementary material~\cite{abelli2024watsurvey}.

\begin{scriptsize}
\begin{longtable}{m{1.5cm}|m{1.3cm}|m{5.5cm}|m{2.7cm}|c}
\caption{A comprehensive overview of tools.} 
\label{TAB:wat_tools} \\
\hline
\textbf{Tool Name} & \textbf{Tool Type}
  & \textbf{Overview} & \textbf{Resource Link} & \textbf{Reference} \\ 
\hline
\endfirsthead

\caption{(Continued).}\\
\hline
\textbf{Tool Name} & \textbf{Tool Type}
  & \textbf{Overview} & \textbf{Resource Link} & \textbf{Reference} \\ 
\hline
\endhead
  Selenium & Open-source 
  & It is open source, cross-platform, and supports multiple languages and browsers.  
  & \url{https://www.selenium.dev/}  
  \newline \url{https://github.com/SeleniumHQ}
  &~\cite{aditya2023ensemble}\\\hline
  ZAP & Open-source
  & It is an open-source web application scanner that excels at finding common web application vulnerabilities.  
  & \url{https://www.zaproxy.org/}  
  \newline \url{https://github.com/zaproxy}
  &~\cite{amankwah2020automated}\\\hline
  JMeter & Open-source 
  & It is an open-source tool for load testing and performance measurement, especially for web applications. 
  It allows the simulation of a large number of users and provides detailed reports.  
  & \url{https://jmeter.apache.org/}  
  \newline \url{https://github.com/apache/jmeter}
  &~\cite{cai2014analysis}\\\hline
  Arachni & Open-source
  & It is a web vulnerability scanner designed to automatically detect and assess security vulnerabilities in web applications.
  & \url{https://github.com/Arachni/arachni}
  &~\cite{roman2018enlargement}\\\hline
  Crawljax & Open-source 
  &It is a tool for automatically crawling and testing web applications that explores the GUI, creates state flow diagrams, and generates test cases based on interactions. 
  & \url{https://github.com/crawljax/crawljax}
  &~\cite{sunman2022automated}\\\hline
  Cucumber & Open-source 
  &It is a tool that automatically runs business case-driven tests, ensuring all scenarios are tested per the acceptance criteria. 
  & \url{https://github.com/cucumber/cucumber}
  &~\cite{rathod2015automatic}\\\hline
  Nikto & Open-source
  & It is an open-source command-line tool that focuses on rapidly scanning web servers and websites for configuration errors, known security vulnerabilities, and missing XSS-protection headers.
  & \url{https://cirt.net/Nikto2}
  &~\cite{devi2020testing}\\\hline
  Puppeteer & Open-source
  & It automates web interactions to perform testing.  
  & \url{https://github.com/puppeteer/puppeteer}
  &~\cite{chen2021improving}\\\hline
  Cypress & Open-source
  & It is a JavaScript-based tool for automated end-to-end testing of modern web applications. 
  & \url{https://github.com/cypress-io/cypress}
  &~\cite{mollah2023user}\\\hline
  Axiom & Open-source
  & It is a dynamic infrastructure framework for distributed computing and testing that automates distributed penetration testing across cloud environments.
  & \url{https://github.com/pry0cc/axiom}
  &~\cite{prisadi2023implementation}\\\hline
  Beautiful Soup & Open-source
  & It can be used to extract data from HTML and XML files.  
  & \url{https://www.crummy.com/software/BeautifulSoup}  
  \newline \url{https://git.launchpad.net/beautifulsoup}
  &~\cite{paul2018approach}\\\hline
  Commix & Open-source
  &It is a tool that automatically detects and exploits command injection vulnerabilities in web applications, including classic, blind, and other types. 
  & \url{https://github.com/commixproject/commix}
  &~\cite{stasinopoulos2019commix}\\\hline
  Katalon Studio & Commercial (Free) 
  & It is easy to install and use, has powerful built-in reporting capabilities, and supports record and playback.  
  & \url{https://katalon.com/}
  &~\cite{nguyen2021generating}\\\hline 
  Cytestion & Open-source
  & It is a web-based tool for automated exploration and testing of GUIs. 
  & \url{https://gitlab.com/lsi-ufcg/cytestion/cytestion}
  &~\cite{moura2023cytestion}\\\hline
  Lazyrecon & Open-source 
  &It is an automated vulnerability-scanning and data-collection framework.  
  & \url{https://github.com/nahamsec/lazyrecon/blob/master/lazyrecon.sh}  &~\cite{shaji2021assessing}\\\hline
  Jena API & Open-source
  &It is an open-source Java framework for building Semantic Web and Linked Data applications that supports SPARQL queries.
  & \url{https://jena.apache.org/}
  &~\cite{durai2021novel}\\\hline
  Binwalk & Open-source
  & It is a firmware analysis tool suitable for most firmware unpacking.
  & \url{https://github.com/ReFirmLabs/binwalk}
  &~\cite{li2023facilitating}\\\hline
  jÄk & Open-source
  &It is a web application crawler and scanner tool that uses dynamic JavaScript code analysis.
  & \url{https://github.com/ConstantinT/jAEk}
  &~\cite{pellegrino2015jak}\\\hline
  Skipfish & Open-source
  &It is an efficient web application security vulnerability scanning tool that quickly detects common web application security issues.  
  & \url{https://code.google.com/archive/p/skipfish/}
  &~\cite{touseef2019analysis}\\\hline
  Sqlmap & Open-source
  & It is a tool for detecting and exploiting SQL injection vulnerabilities.
  & \url{https://sqlmap.org/}
  &~\cite{prisadi2023implementation}\\\hline
  Scrapy & Open-source 
  & It is a tool for quickly crawling HTML-based static web page content. 
  & \url{https://github.com/scrapy/scrapy}
  &~\cite{rennhard2022automating}\\\hline
  Metasploit & Open-source
  & It is an open-source framework for penetration testing and vulnerability exploitation.  
  & \url{https://www.metasploit.com/}  
  \newline \url{https://github.com/rapid7/metasploit-framework}
  &~\cite{jana2019appmine}\\\hline
  OWASP AppSensor & Open-source
  &It is a tool that detects suspicious application-layer behavior and triggers events based on predefined sensor conditions. 
  & \url{https://owasp.org/www-project-appsensor/}
  &~\cite{shahrivar2023detecting}\\\hline
  HTMLCS & Open-source
  & It is a tool for checking whether HTML code complies with WCAG standards. 
  & \url{http://squizlabs.github.io/HTML_CodeSniffer/}
  &~\cite{campoverde2021process}\\\hline
  GPTFuzzer & Open-source
  & It is a tool for generating web application firewall (WAF) bypass payloads, testing, and discovering common web attacks such as SQL injection, cross-site scripting (XSS), and remote command execution (RCE) vulnerabilities.
  & \url{https://github.com/hongliangliang/gptfuzzer}
  &~\cite{liang2023generative}\\\hline
  SideeX & Open-source
  &It is a tool that provides cross-browser web UI automated testing, supports recording and playback, and optimizes the test process through an automatic waiting mechanism.  
  & \url{http://sideex.org}
  &~\cite{lee2018test}\\\hline
  CICFlowMeter & Open-source
  &It is a tool that provides network flow signatures for intrusion detection and classification of malicious behavior. 
  & \url{https://github.com/ahlashkari/CICFlowMeter}
  &~\cite{rajic2023early}\\\hline
  WebPageTest & Commercial (Free)  
  &It is used to analyze web page performance, providing details such as page load time and resource load times.  
  & \url{https://www.webpagetest.org/}
  &~\cite{ali2015case}\\\hline
  WebTest & Open-source
  &It is a tool that supports visualization of output coverage, helps testers understand which parts of the output are covered or uncovered, and optimizes test cases. 
  & \url{https://github.com/git1997/VarAnalysis}
  &~\cite{nguyen2019exploring}\\\hline
  GTmetrix & Commercial (Free) 
  &It is used to measure web page loading speed and performance and to provide detailed optimization suggestions.  
  & \url{https://gtmetrix.com/}
  &~\cite{ali2015case}\\\hline
  SikuliX & Open-source
  &It is an image-based testing tool that can control user input methods for interacting with GUI elements, suitable for both web and non-web applications.
  & \url{http://sikulix.com}
  &~\cite{macchi2021image}\\\hline
  Hakrawler & Open-source
  & It is a web crawler tool that collects web links and identifies injection points.
  & \url{https://github.com/hakluke/hakrawler}
  &~\cite{prisadi2023implementation}\\\hline
  Axe Studio & Commercial (Free) 
  & It is a tool that focuses on automated detection of web accessibility issues. 
  & \url{https://www.deque.com/axe/}
  &~\cite{campoverde2021process}\\\hline
  Cornipickle & Open-source
  & It is a declarative testing tool that allows developers to define layout errors using logical expressions and compare states of different snapshots. 
  & \url{https://github.com/liflab/cornipickle}
  &~\cite{beroual2020detecting}\\\hline
  FitNesse & Open-source
  & It can automate user scenario testing of web applications for acceptance testing and system verification.  
  & \url{http://www.fitnesse.org}  
  \newline \url{https://github.com/unclebob/fitnesse}
  &~\cite{otaduy2017user}\\\hline
  Wapiti & Open-source
  & It is an open-source web application scanner that focuses on injection attacks and authentication-protected page scanning.  
  & \url{https://wapiti-scanner.github.io}  
  \newline \url{https://github.com/wapiti-scanner/wapiti}
  &~\cite{anagandula2020analysis}\\\hline
  Ultralytics YOLOv8 & Open-source
  & It can perform object detection on UI elements in screenshots, providing fine-grained analysis.  
  & \url{https://github.com/ultralytics/ultralytics}
  &~\cite{aditya2023ensemble}\\\hline
  SoapUI & Open-source
  & It supports multiple protocols (SOAP, HTTP, JDBC, etc.), provides statistics and historical statistics charts, and supports cross-platform load testing.  
  & \url{https://www.soapui.org/}  
  \newline \url{https://github.com/SmartBear/soapui}
  &~\cite{tiwari2023analytical}\\\hline
  Sysdig & Open-source
  & It is a monitoring tool that collects detailed system call data for applications running in Docker containers.  
  & \url{https://sysdig.com/}  
  \newline \url{https://github.com/draios/sysdig}
  &~\cite{jana2019appmine}\\\hline
  Sublist3r & Open-source 
  & It is an automated subdomain enumeration tool that supports multiple DNS record lookups.  
  & \url{https://github.com/aboul3la/Sublist3r}
  &~\cite{shaji2021assessing}\\\hline
 GitLeaks & Open-source 
  & It is a tool for finding leaks of sensitive information in GitHub repositories.  
  & \url{https://github.com/zricethezav/gitleaks}
  &~\cite{shaji2021assessing}\\\hline
\end{longtable}
\end{scriptsize}

Open-source tools like Selenium~\cite{dawei2016Web}, JMeter~\cite{cai2014analysis}, and ZAP~\cite{touseef2019analysis} provide cross-platform compatibility, support multiple programming languages, and have received strong community support. 
They benefit from scalability, flexibility, and collaborative development through platforms like GitHub, where contributors can provide updates and improvements.
The reviewed literature further shows that browser automation frameworks (such as Selenium, Playwright, and WebdriverIO) have been integrated into LLM-assisted workflows for test generation and wait management.
Such integration can improve test-authoring efficiency while preserving the repeatability of test execution~\cite{aryotejo2025comparative,olianas2025leveraging}.

In contrast, commercial tools such as Katalon Studio~\cite{nguyen2021generating} and GTmetrix~\cite{ali2015case} often prioritize usability and integrated reporting, making them more suitable for testers and organizations that value refined interfaces and out-of-the-box functionality. 
However, limitations in the free versions may make them less suitable for certain users.

Nearly all of the reviewed tools are easily accessible through official websites or open-source platforms.
This enables a smooth integration into a wide range of testing environments. 
Comprehensive documentation and active community support further simplify adoption and customization. 

The reviewed literature also demonstrates how complementary WAT tools and techniques can be combined across different testing activities.
In security testing, tools such as sqlmap, Metasploit, and Burp Suite can be incorporated into coordinated workflows or supplemented with manual verification to broaden vulnerability coverage and reduce false positives~\cite{samgir2024automated,osman2025comparative,juliadi2025web,nana2025bf}.
Accessibility testing similarly combines automated checks with assessments of WCAG conformance, reflow behavior, color contrast, and screen-reader interaction across desktop and mobile environments~\cite{manu2025enhancing,sivayoganathan2025evaluating,kalac2025digital,he2025enhancing}.
Performance-oriented tools such as Lighthouse and PageSpeed Insights can also be used alongside accessibility audits, allowing performance and accessibility dimensions of user-facing quality to be evaluated together~\cite{giambartolomei2024penetration,sivayoganathan2025evaluating}.

The tools offer diverse capabilities, ranging from identifying security vulnerabilities with tools such as Wapiti~\cite{anagandula2020analysis} and Skipfish~\cite{touseef2019analysis} to improving page performance with GTmetrix~\cite{ali2015case} and WebPageTest~\cite{ali2015case}.
Their wide availability and applicability underscore their role in advancing WAT practices.

\section{ANSWER TO RQ6: Challenges and Future Work}
\label{SEC:rq6}

This section addresses RQ6 by identifying unresolved challenges in WAT and outlining directions for future research.
We first examine the principal challenges that continue to affect the field and then discuss research directions that may help address these limitations and advance WAT practice.

The growing adoption of LLMs, LVLMs, and agent-based systems has broadened the scope of WAT to include test-script generation, GUI testing, fuzzing, penetration testing, and user-behavior simulation~\cite{leotta2024ai,wang2024leveraging,liang2023generative,wu2025autopt,lu2025uxagent}.
Although these approaches create new opportunities for WAT, they also introduce challenges, particularly in ensuring context awareness, robustness, and reliability when testing dynamic and complex web applications.

\subsection{Challenges}
Despite notable advances in recent years, some significant challenges to WAT remain, preventing the full realization of its potential. 
Here we discuss several key challenges:

\begin{itemize}
    \item
    As web applications become increasingly complex, automated testing continues to face scalability challenges, particularly when exploring large and dynamically changing state spaces.
    Existing techniques still have difficulty handling large-scale testing processes and dynamic web content~\cite{pellegrino2015jak}.
    Reinforcement-learning-based and agent-based approaches can improve web exploration, but single-agent methods may struggle to explore large state spaces efficiently~\cite{fan2024can}.
    Cooperative multi-agent approaches may extend exploration and improve coverage, although they can introduce additional communication, coordination, and computational overhead~\cite{wu2025autopt,juliadi2025web}.
    Efficiently scaling automated testing to complex and dynamic web applications therefore remains an unresolved challenge.

    \item
    Maintaining reliable test scripts for dynamic web applications is another significant challenge.
    Frequent changes to UI components and DOM structures can invalidate existing test cases and increase maintenance effort~\cite{ryou2018automatic}.
    LLMs and other AI-based approaches can generate or adapt web test scripts, potentially reducing development effort.
    Nevertheless, generated locators may be inaccurate, and the resulting code may still require manual refinement~\cite{leotta2024ai}.
    Dynamic DOM updates, asynchronous operations, and rendering delays can further undermine test reliability.
    Although synthetic-data and scenario-generation techniques may improve this process, fixed prompting strategies and rule-based element matching remain limited when applied to dynamic or JavaScript-intensive applications~\cite{manivannan2025ai,olianas2025leveraging,wang2024leveraging}.
    Improving the ability of AI-based testing systems to incorporate and respond to the current application state is therefore an important research challenge.

    \item
    Balancing test coverage against execution efficiency remains difficult in web testing.
    Virtual DOM coverage techniques have demonstrated the potential to improve testing efficiency for dynamic web applications~\cite{zou2014virtual}, while hyper-heuristic approaches can reduce test-case redundancy~\cite{balera2022multiperspective}.
    AI-based testing, however, introduces additional trade-offs among exploration breadth, execution cost, and feedback quality.
    These trade-offs are also evident in security testing, where LLM-based techniques can generate fuzzing inputs and security payloads.
    Although such techniques can improve payload generation, their effectiveness depends on factors such as payload-dataset quality, grammar design, reward signals, and search strategies~\cite{liang2023generative,chen2024webfuzzauto,yang2024llmsqli}.
    Limited grammar coverage and sparse feedback can further constrain the testing process.
    Achieving comprehensive coverage without substantially increasing testing cost therefore remains an open challenge.

    \item
    Failure detection and result verification remain major bottlenecks in WAT.
    Model-based techniques can assist with failure identification, yet false positives and false negatives continue to affect their reliability~\cite{alkofahi2022discovering}.
    Agent-based systems extend automation to more complex activities, including web exploration and penetration testing.
    Frameworks such as AutoPT and collaborative multi-agent approaches demonstrate the potential to automate multiple stages of these activities~\cite{wu2025autopt,juliadi2025web}.
    However, many existing approaches still rely on structured control flows, sequential execution, or narrowly defined tasks.
    Reliable failure verification, management of unexpected application states, result reporting, and recovery from unsuccessful actions all require further improvement.
    Supporting a dependable end-to-end testing workflow therefore remains a central challenge for agent-based WAT.

    \item
    The absence of standardized metrics for evaluating testing effectiveness remains a significant challenge.
    The use of heterogeneous evaluation measures across studies makes direct comparisons difficult.
    Unified evaluation frameworks and consistent benchmarks are needed to assess test coverage and efficiency~\cite{sonmez2021holistic}.
    Although initiatives such as WebEV help standardize certain aspects of test-behavior analysis, further work is required~\cite{fuad2023webev}.
    Evaluation becomes even more complex for AI-based testing, for which studies report measures such as script-development time, explored states, unique actions, test coverage, bypass rate, task success, and manual-review results~\cite{leotta2024ai,wang2024leveraging,fan2024can,lu2025uxagent,wu2025autopt}.
    More consistent evaluation methods are therefore needed to assess effectiveness, efficiency, robustness, and the degree of human intervention required.

    \item
    Human factors remain central to the application of AI-based approaches in WAT.
    LLM-based agents can simulate users and evaluate web interfaces, and systems such as UXAgent indicate that simulated users can provide useful feedback during usability evaluation~\cite{lu2025uxagent}.
    However, generated behavior cannot yet be assumed to represent real human behavior faithfully and may reflect biases inherited from the underlying models.
    Similar concerns apply to AI-assisted web testing, in which generated scripts, actions, and results may still require human verification.
    A key challenge is therefore determining when AI-generated testing results are sufficiently reliable and when additional human validation remains necessary.
\end{itemize}

\subsection{Future Work}
Potential directions for future WAT research include advancing context-aware testing, strengthening agent-based workflows, improving security-payload generation, standardizing evaluation practices, incorporating appropriate human oversight, and enhancing deployment support.

\begin{itemize}
    \item
    Future research should develop more context-aware testing techniques for dynamic web applications.
    Combining DOM information, screenshots, and LVLM reasoning could improve input generation, element selection, and interaction planning based on the current page state~\cite{wang2024leveraging}.
    Self-healing locator strategies, adaptive synchronization, and explicit wait mechanisms should also be investigated to improve the handling of dynamic web interfaces~\cite{emektar2025ai,olianas2025leveraging}.

    \item
    Agent-based WAT approaches require stronger mechanisms for control and verification.
    Future systems should incorporate explicit result validation, fallback and recovery mechanisms, and improved state management to support more open-ended applications and reasoning across multiple pages or targets~\cite{wu2025autopt,juliadi2025web,fan2024can}.
    These improvements could enable agent-based methods to support additional stages of the testing lifecycle
    ---
    from exploration and test generation to failure verification and reporting.

    \item
    Improving security-payload generation represents another important research direction.
    Transformer-based generation, reward modeling, and KL regularization have the potential to improve payload discovery, but current approaches may remain dependent on manually designed grammars or fixed payload datasets~\cite{liang2023generative}.
    Future work should reduce this dependence and evaluate such techniques across a broader range of vulnerability classes and security-testing settings~\cite{chen2024webfuzzauto,yang2024llmsqli}.

    \item
    Standardized protocols for evaluating WAT effectiveness are also needed.
    Future studies should consistently report measures of test quality, execution cost, coverage, robustness, and required human intervention~\cite{leotta2024ai,wang2024leveraging,wu2025autopt,lu2025uxagent}.
    Larger datasets, more realistic web applications, and clearly specified benchmark protocols would further support meaningful comparisons among testing approaches.

    \item
    Human oversight should remain an integral component of AI-assisted web testing.
    Although AI-based user simulation can provide useful feedback, it should complement rather than replace evaluation involving real users~\cite{lu2025uxagent}.
    Future research should also examine bias, privacy, and interpretability and establish when expert review is required.

    \item
    Future WAT tools should offer greater robustness and deployment readiness.
    Research should investigate CI/CD integration, reporting dashboards, and broader deployment settings, including mobile and hybrid environments~\cite{emektar2025ai,olianas2025leveraging,lu2025uxagent}.
    These tools should also maintain reliable operation despite layout changes, timing variability, and other common sources of web-automation failure.
\end{itemize}

\section{Conclusions
\label{SEC:conclusions}}

This review has provided a comprehensive analysis of advances in \textit{Web Application Testing} (WAT) over the twelve-year period from January 1, 2014, to December 31, 2025.
A total of 464 WAT papers were examined using a rigorous systematic review methodology.
Five major academic databases were searched using targeted keywords, and the retrieved papers were screened against predefined selection criteria to ensure their relevance to the study.

The review was structured around six research questions covering the evolution of WAT research, test-case generation and execution, evaluation metrics, testing tools, and unresolved challenges.
The findings indicate that WAT has remained an active research area, with a notable increase in publications during 2024 and 2025.
Its scope has also expanded beyond traditional functional and regression testing to encompass security, accessibility, usability, performance, privacy, GUI exploration, and AI-assisted testing.

The review examined a diverse range of test-case generation methods, including security-based, model-based, data-driven, AI-based, optimization-based, and exploration-based techniques.
Among these, security-oriented test-case generation emerged as the most prominent category.
With respect to test execution, automated execution was the dominant approach, while hybrid execution remained important in scenarios requiring human judgment or domain expertise.
The review also considered local, remote, and cloud-based execution environments.
Metrics for evaluating WAT effectiveness and efficiency included detection capability, coverage, test-case quality, execution time, resource utilization, and cost.
In addition, WAT tools were categorized according to their functions and accessibility, providing a practical overview for both researchers and practitioners.

Several persistent challenges were also identified, including dynamic content, asynchronous behavior, complex GUI states, changing DOM structures, large input spaces, and evolving security threats.
LLM-, LVLM-, and agent-based techniques have extended WAT capabilities in areas such as test generation, GUI exploration, security testing, and user-behavior simulation.
However, these techniques also introduce concerns regarding robustness, model hallucinations, reproducibility, result verification, computational cost, and human oversight.
Future research should therefore prioritize robust and context-aware automation, stronger evaluation and verification methods, scalable and reproducible testing practices, and the responsible application of AI-assisted techniques.

Overall, this review consolidates the major developments, trends, and unresolved challenges in WAT from 2014 to 2025.
It provides a useful resource for researchers and practitioners seeking to improve the quality, reliability, accessibility, and security of modern web applications.

\section*{Acknowledgment}
This work in this paper was partly supported by the Science and Technology Development Fund of Macau, Macao SAR (Grant Nos. 0069/2025/RIB2 and 0021/2023/RIA1).

\bibliographystyle{ACM-Reference-Format}
\bibliography{acm-ref}

\end{document}